\documentclass[aps, prd, amsmath, floats, floatfix, twocolumn, nofootinbib, superscriptaddress]{revtex4-1}
\usepackage[pdftex]{graphicx}
\usepackage{color}
\usepackage{latexsym}
\usepackage[colorlinks]{hyperref}
\usepackage{amsthm,amsmath,amssymb}
\usepackage{mathrsfs}
\usepackage{makecell}

\newcommand{\be}{\begin{eqnarray}}
\newcommand{\ee}{\end{eqnarray}}

\begin{document}

\title{About the ability of agnostic X-ray tests of the Kerr Hypothesis\\to discover new physics}

\author{Shuaitongze~Zhao}
\affiliation{Center for Field Theory and Particle Physics and Department of Physics, Fudan University, 200438 Shanghai, China}

\author{Shafqat~Riaz}
\affiliation{Theoretical Astrophysics, Eberhard-Karls Universit\"at T\"ubingen, D-72076 T\"ubingen, Germany}

\author{Cosimo~Bambi}
\email[Corresponding author: ]{bambi@fudan.edu.cn}
\affiliation{Center for Field Theory and Particle Physics and Department of Physics, Fudan University, 200438 Shanghai, China}
\affiliation{School of Natural Sciences and Humanities, New Uzbekistan University, Tashkent 100007, Uzbekistan}

\begin{abstract}
In the past decade, we have seen an unprecedented progress in our ability of testing general relativity in the strong field regime with black hole observations. Most studies have focused on the so-called tests of the Kerr hypothesis: they have tried to verify whether the spacetime geometry around black holes is described by the Kerr solution as expected in general relativity. One can follow either a theory-specific analysis or an agnostic approach. Each strategy has its advantages and disadvantages. In this work, we study the ability of agnostic X-ray tests of the Kerr hypothesis to discover new physics. We simulate X-ray observations of bright Galactic black holes of specific theories of gravity and we analyze the simulated data with a reflection model employing the correct theory of gravity and another reflection model for agnostic tests of the Kerr hypothesis. Our results suggest that agnostic X-ray tests are valid tools to discover new physics, but their constraining power may be lower than a theory-specific analysis.   
\end{abstract}

\maketitle


\section{Introduction}

Einstein's theory of general relativity is one of the pillars of modern physics and our current standard framework for the description of gravitational interactions and of the chrono-geometrical structure of the spacetime. For decades, the theory has been extensively tested in the weak field regime with experiments in the Solar System and accurate radio observations of binary pulsars~\cite{Will:2014kxa}. Black holes are ideal laboratories for testing general relativity in the strong field regime~\cite{Bambi:2015kza,Bambi:2017khi,Yagi:2016jml,Barack:2018yly}. The past 8~years have seen unprecedented progress in our ability of testing general relativity in the strong field regime with black hole observations, in particular with gravitational waves~\cite{LIGOScientific:2016lio,Yunes:2016jcc,LIGOScientific:2020tif,Shashank:2021giy}, X-ray data~\cite{Cao:2017kdq,Tripathi:2018lhx,Tripathi:2020dni,Tripathi:2020yts}, and black hole imaging~\cite{Bambi:2019tjh,EventHorizonTelescope:2020qrl,EventHorizonTelescope:2022xqj,Vagnozzi:2022moj}.

In 4-dimensional general relativity, uncharged black holes are described by the Kerr solution~\cite{Kerr:1963ud}, which is completely characterized by the black hole mass, $M$, and the black hole spin angular momentum, $J$. Deviations from the Kerr metric due to accretion disks, nearby stars, or some initial non-vanishing electric charge are normally completely negligible (see, for instance, Refs.~\cite{Bambi:2014koa,Bambi:2008hp}). On the other hand, macroscopic deviations from the Kerr geometry are possible in the presence of new physics (see, for instance, Refs.~\cite{Dvali:2011aa,Herdeiro:2014goa,Giddings:2014ova}). Tests of the Kerr hypothesis can thus be seen as the strong field counterpart of the tests of the Schwarzschild metric in the weak field limit with Solar System experiments.

Unlike Solar System experiments, there is no common consensus on how to test the Kerr hypothesis with black holes. As in many other contexts, we can think about two possible strategies, which are normally referred to as {\it top-down} (or theory-specific) approach and {\it bottom-up} (or agnostic) approach.

The theory-specific strategy is the most logical one: we want to test general relativity against some other specific theory of gravity. In such a case, we can analyze some black hole observations with a theoretical model in general relativity and with the same theoretical model in the other theory of gravity. With the use of some statistical tool we can compare the two results and see if one of the two models can explain the data better and if we can rule out the other one. The main technical problem with top-down tests of general relativity is that we need to have a good understanding of the predictions of the other theory of gravity, which is normally not the case. If we want to test the Kerr hypothesis with X-ray data, we need to know the rotating black hole solution of the other theory of gravity. However, for most theories of gravity we only know their non-rotating black hole solution, or at best the black hole solution in the slow rotation approximation, while the complete rotating solution is often unknown. This is just because it is much easier to find spherically symmetric solutions than axially symmetric ones and it was true even in general relativity, where the non-rotating Schwarzschild solution was found by Schwarzschild in 1916, immediately after Einstein had presented his theory, while the rotating Kerr solution was found by Kerr only in 1963. Another unpleasant aspect of these theory-specific tests is that there are many theories of gravity beyond general relativity, but none of them is more promising than the other ones, so we should repeat a test for every theory of gravity, which would be extremely time-consuming.

Agnostic tests of the Kerr hypothesis rely on parametric black hole spacetimes in which a finite or infinite set of {\it deformation parameters} are introduced to quantify possible deviations from the Kerr solution. In general, these metrics are not solutions of specific field equations and they are instead obtained by deforming the Kerr metric under certain conditions (e.g. requiring the existence of a Carter-like constant, that event horizons and Killing horizons coincide, etc.). Ideally, one would like to have a metric general enough to be able to describe any black hole solution from any consistent theory of gravity. In practice, this is not easy to achieve. Unlike in the case of Solar System tests, it is not possible to perform an expansion in $M/r$ (where $M$ is the black hole mass and $r$ some radial coordinate), because $M/r$ is not a small parameter when we probe the strong gravity region around black holes. There is no natural way to deform the Kerr metric and deformations often lead to pathological properties like naked singularities, regions with closed time-like curves, etc. Despite these issues, agnostic tests of the Kerr hypothesis have been quite popular in the past years because they can be seen as null experiments: we expect that deviations from the Kerr solution vanish and we want to check if observations can confirm it.

In the present work, we want to figure out if the agnostic X-ray tests of the Kerr hypothesis are valid tools for discovering new physics. This is quite a natural question, as this is often the only strategy that we can follow because $i)$~we do not know the actual spacetime geometry around black holes (and despite the large number of theories of gravity beyond general relativity proposed in the literature, we may not yet know the correct one), and $ii)$~we do not know the rotating black hole solutions for most theories of gravity beyond general relativity. To address our question, we adopt the following approach. We consider some non-Kerr black hole solutions from specific theories of gravity and we simulate the observation of a bright Galactic black hole with strong reflection features with \textsl{NuSTAR}~\cite{NuSTAR:2013yza}, which is currently the most suitable X-ray mission for testing general relativity with black hole X-ray data. The reflection features of the simulated spectra are analyzed both with a reflection model employing the correct background metric and with a reflection model employing a parametric black hole spacetime. We consider two specific examples of black hole spacetimes from theories beyond general relativity: black holes in conformal gravity~\cite{Bambi:2016wdn} and black holes in Einstein-Maxwell-dilaton-axion gravity~\cite{Sen:1992ua}. For the agnostic tests, our reflection model employs the Johannsen spacetime~\cite{Johannsen:2013szh}, which has been extensively used to test the Kerr hypothesis with X-ray data (see~\cite{Zhang:2021ymo,Tripathi:2021rqs,Bambi:2022dtw} and references therein), black hole imaging~\cite{EventHorizonTelescope:2020qrl,EventHorizonTelescope:2022xqj}, and gravitational waves~\cite{Psaltis:2020ctj}.

The manuscript is organized as follows. In Section~\ref{s-xrs}, we briefly review X-ray reflection spectroscopy, which is the leading technique to test the Kerr hypothesis with black hole X-ray data. In Section~\ref{s-sim}, we present our simulations and spectral analyses: in Subsection~\ref{ss-1} we consider black holes in conformal gravity and in Subsection~\ref{ss-2} black holes in Einstein-Maxwell-dilaton-axion gravity. In both studies, we fit the simulated data first with a model employing the correct background metric and then with a model adopting the Johannsen spacetime with non-vanishing deformation parameter $\alpha_{13}$. Discussion and conclusions are reported in Section~\ref{s-d}. The metrics of black holes in conformal gravity, of black holes in Einstein-Maxwell-dilaton-axion gravity, and of the Johannsen spacetime are reported, respectively, in Appendix~\ref{aaa-1}, \ref{aaa-2}, and \ref{aaa-3}.


\section{X-ray reflection spectroscopy}\label{s-xrs}

\begin{figure}[t]
\begin{center}
\includegraphics[width=0.95\linewidth]{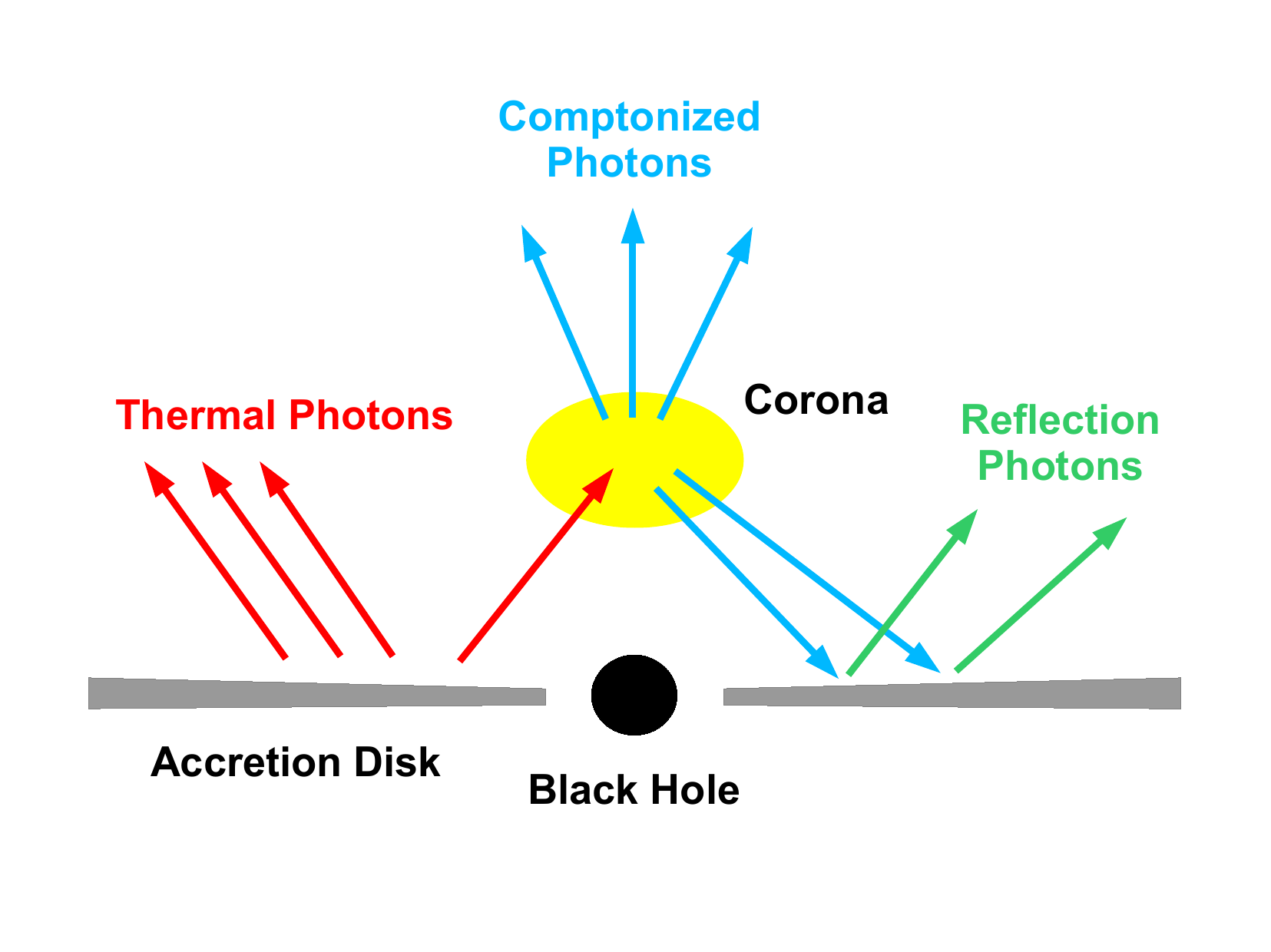}
\end{center}
\vspace{-1.2cm}
\caption{Disk-corona model. See the text for the description of the system. Figure from Ref.~\cite{Bambi:2021chr} under the terms of the Creative Commons Attribution 4.0 International License. \label{f-corona}}
\end{figure}

X-ray tests of the Kerr hypothesis require the astrophysical system shown in Fig.~\ref{f-corona}, which is normally referred to as the disk-corona model (for more details on the disk-corona model, see \cite{Bambi:2020jpe} and references therein). The black hole can be either a stellar-mass black hole in an X-ray binary or a supermassive black hole in an active galactic nucleus. The key-point is that the black hole is accreting from a geometrically thin and optically thick accretion disk. Every point on the surface of the disk emits a blackbody-like spectrum and the whole disk has a multi-temperature blackbody-like spectrum. The disk is ``cold'' because it can efficiently emit radiation. Since the maximum temperature scales as $M^{-1/4}$~\cite{Bambi:2012tg}, the thermal spectrum of the disk normally peaks in the soft X-ray band for stellar-mass black holes and in the UV band for supermassive black holes. Thermal photons of the disk can inverse Compton scatter off free electrons in the ``corona'', which is some ``hot'' (with respect to the disk) plasma near the black hole and the inner part of the accretion disk. The spectrum of the Comptonized photons can be usually approximated well by a power law component with an exponential high-energy cutoff. A fraction of the Comptonized photons can illuminate the disk: here we have Compton scattering and absorption followed by fluorescent emission, which generate a reflection spectrum.

In the rest-frame of the plasma of the disk, the reflection spectrum is characterized by narrow fluorescent emission lines in the soft X-ray band and by a Compton hump with a peak around 20-30~keV~\cite{Ross:2005dm,Garcia:2010iz}. The most prominent line in the reflection spectrum is usually the iron K$\alpha$ complex, which is a narrow line at 6.4~keV in the case of neutral or weakly ionized iron atoms and shifts up to 6.97~keV in the case of H-like iron ions. The relativistic reflection spectrum, which is the reflection spectrum of the whole disk as seen by a distant observer, is blurred due to relativistic effects (Doppler boosting and gravitational redshift). In the presence of high-quality data and employing the correct astrophysical model, the analysis of these relativistically blurred reflection features is a powerful tool to probe the strong gravity region around black holes, measure black hole spins, and even test Einstein's theory of general relativity in the strong field regime~\cite{Bambi:2020jpe}.


\section{Simulations and spectral analysis}\label{s-sim}

In this work, we use the reflection model {\tt relxill\_nk}~\cite{Bambi:2016sac,Abdikamalov:2019yrr,Abdikamalov:2020oci}, which is an extension of the {\tt relxill} model~\cite{Dauser:2013xv,Garcia:2013oma,Garcia:2013lxa} to non-Kerr spacetimes and is public on GitHub~\cite{Bambi:2023czx}\footnote{\url{https://github.com/ABHModels}}. In particular, we use two versions of {\tt relxill\_nk}: the version with black holes in conformal gravity~\cite{Zhou:2018bxk,Zhou:2019hqk} and that with black holes in Einstein-Maxwell-dilaton-axion gravity~\cite{Tripathi:2021rwb}.

We simulate 30~ks observations of these black holes with \textsl{NuSTAR}, assuming that the sources are bright Galactic black holes (we set the flux of every source to $10^{-8}$~erg~cm$^{-2}$~s$^{-1}$ in the energy band 1-10~keV). In XSPEC language, the simulated model is simply

\vspace{0.2cm}

{\tt tbabs $\times$ relxill\_nk} 

\vspace{0.2cm}

\noindent where {\tt tbabs} takes the Galactic absorption into account~\cite{Wilms:2000ez} and {\tt relxill\_nk} describes both the continuum from the corona and the relativistically blurred reflection spectrum from the disk. {\tt tbabs} has only one parameter, the hydrogen column density $n_{\rm H}$, which is set to $6.74 \times 10^{20}$~cm$^{-2}$ in our simulations. The values of the input parameters in {\tt relxill\_nk} are shown in Tab.~\ref{t-sim}. The values of the black hole spin parameter $a_*$ and of the deformation parameter ($L/M$ for the black holes in conformal gravity and the dilaton parameter $r_2$ for those in Einstein-Maxwell-dilaton-axion gravity) are chosen according to the specific properties of the black hole metric (see below). In all our simulations, we assume that the Comptonized spectrum from the corona is described by a power law with an exponential high-energy cutoff, and we set the photon index to $\Gamma = 1.7$ (which is the typical value of the photon index when a Galactic black hole is in a hard state) and the high-energy cutoff to $E_{\rm cut} = 300$~keV. The reflection fraction $R_{\rm f}$, regulating the relative intensity between the reflection component from the disk and the continuum from the corona, is set to 2. The inner edge of the disk is assumed to be at the innermost stable circular orbit (ISCO), so it is not a free parameter but is determined by the black hole spin and the deformation parameter. We always assume that the inclination angle of the disk with respect to the line of sight of the distant observer, $i$, is $60^\circ$, that the emissivity profile of the reflection spectrum is described by a power law with emissivity index $q = 8$ (i.e. the emissivity of the disk scales as $1/r^8$), that the ionization parameter of the disk is $\xi = 100$~erg~cm~s$^{-1}$ (the actual parameter appearing in {\tt relxill\_nk} is $\log\xi$, so we set $\log\xi = 2$), and that the disk has Solar iron abundance ($A_{\rm Fe} = 1$). A detailed description of the parameters in {\tt relxill\_nk} and of their physical meaning can be found in Ref.~\cite{Bambi:2023czx}. The details of the specific simulations and the corresponding spectra analysis is reported in the next subsections.

\begin{table*}
\renewcommand\arraystretch{1.5}
	\centering
	\vspace{0.5cm}
	\begin{tabular}{lcccccccccc}
		\hline\hline
		Simulation & \hspace{0.3cm} $a_*$ \hspace{0.3cm} & \hspace{0.3cm} $L/M$ \hspace{0.3cm} & \hspace{0.3cm} $r_2$ \hspace{0.3cm} & \hspace{0.3cm} $i$ [deg] \hspace{0.3cm} & \hspace{0.3cm} $q$ \hspace{0.3cm} & \hspace{0.3cm} $\Gamma$ \hspace{0.3cm} & \hspace{0.3cm} $E_{\rm cut}$ [keV] \hspace{0.3cm} & \hspace{0.3cm} $\log\xi$ \hspace{0.3cm} & \hspace{0.3cm} $A_{\rm Fe}$ \hspace{0.3cm} & \hspace{0.3cm} $R_{\rm f}$ \hspace{0.3cm} \\
		\hline\hline
		1 & 0.99 & 0.25 & -- & 60 & 8 & 1.7 & 300 & 2 & 1 & 2 \\
		\hline
		2 & 0.99 & 0.15 & -- & 60 & 8 & 1.7 & 300 & 2 & 1 & 2 \\
		\hline
		3 & 0.2 & -- & 1.4 & 60 & 8 & 1.7 & 300 & 2 & 1 & 2 \\
		\hline
		4 & 0.2 & -- & 1.2 & 60 & 8 & 1.7 & 300 & 2 & 1 & 2 \\
		\hline\hline
	\end{tabular}
	\caption{Summary of the input parameters in Simulations~1-4. $\xi$ in units erg~cm~s$^{-1}$. \label{t-sim}}
\end{table*}


\subsection{Conformal gravity}\label{ss-1}

As our first case study, we consider black holes in conformal gravity. The metric is reported in Appendix~\ref{aaa-1}. The spacetime is characterized by the black hole mass, $M$, the dimensionless spin parameter, $a_* = J/M^2$, and the conformal parameter $L$ ($L \ge 0$). For $L=0$, we recover the Kerr solution of general relativity while for $L \neq 0$ the metric describes the spacetime of a regular black hole without singularities. It is often more convenient to use the quantity $L/M$ (instead of $L$) because it is dimensionless. The implementation of this black hole spacetime in {\tt relxill\_nk} was presented in Ref.~\cite{Zhou:2018bxk}.

To maximize the relativistic effects in the reflection spectrum, the inner edge of the accretion disk should be as close as possible to the black hole and the corona should illuminate well the inner part of the disk. As in the case of Kerr black holes, the ISCO of black holes in conformal gravity is close to the black hole event horizon for high values of $a_*$ and therefore in our simulations we assume $a_* = 0.99$. As discussed above, we set the emissivity index to $q=8$ to meet the requirement that the corona illuminates well the inner part of the accretion disk. In Simulation~1, we set $L/M = 0.25$. In Simulation~2, we have $L/M = 0.15$.

First, we analyze Simulation~1. We fit the simulated spectrum with the model

\vspace{0.2cm}

{\tt constant $\times$ tbabs $\times$ relxill\_nk} 

\begin{figure*}[t]
\centering
\includegraphics[width=0.48\textwidth,trim=2.5cm 1.0cm 2.5cm 0.0cm,clip]{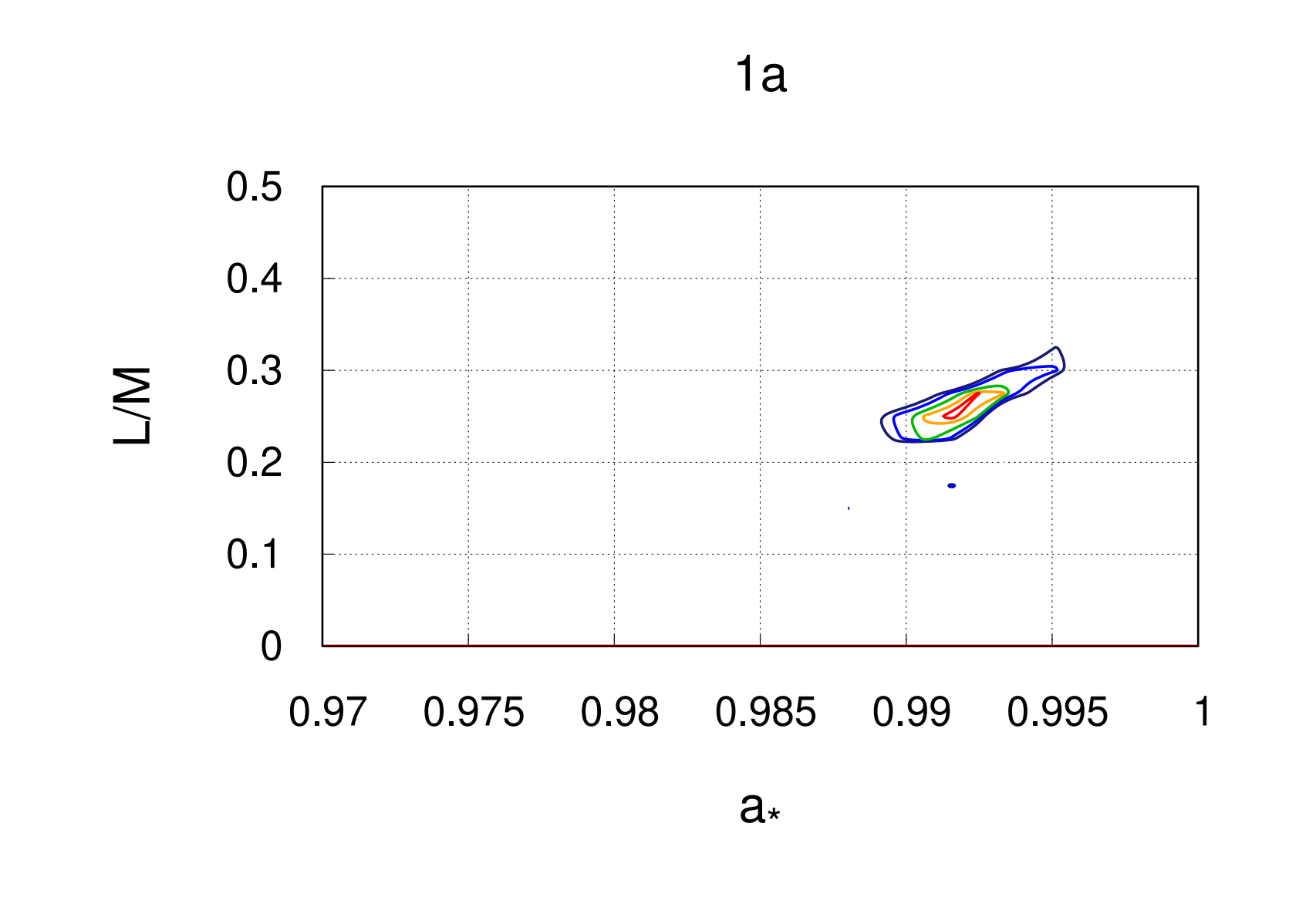} \hspace{0.5cm}
\includegraphics[width=0.48\textwidth,trim=2.5cm 1.0cm 2.5cm 0.0cm,clip]{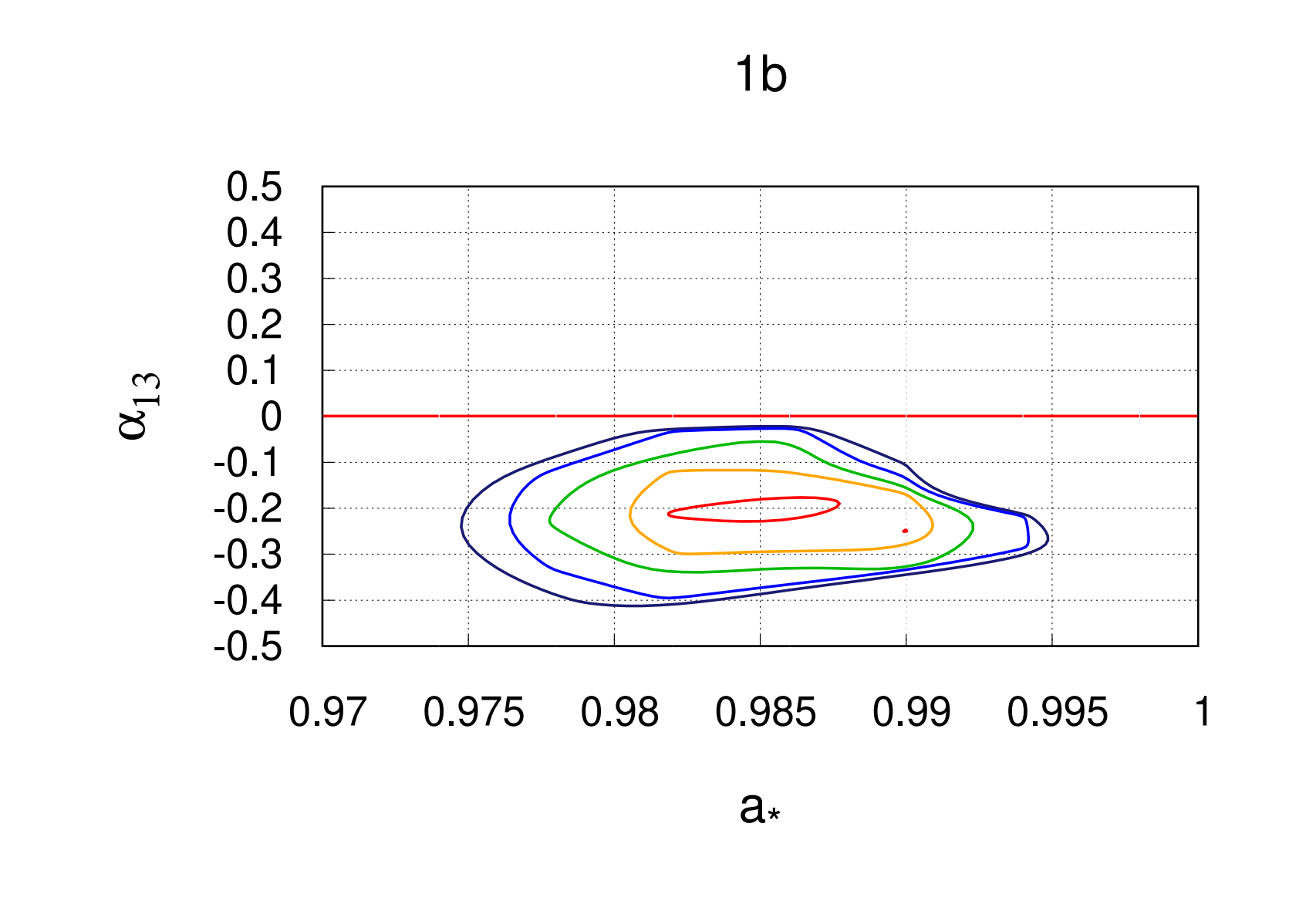} 
\includegraphics[width=0.48\textwidth,trim=2.5cm 2.0cm 2.5cm 0.0cm,clip]{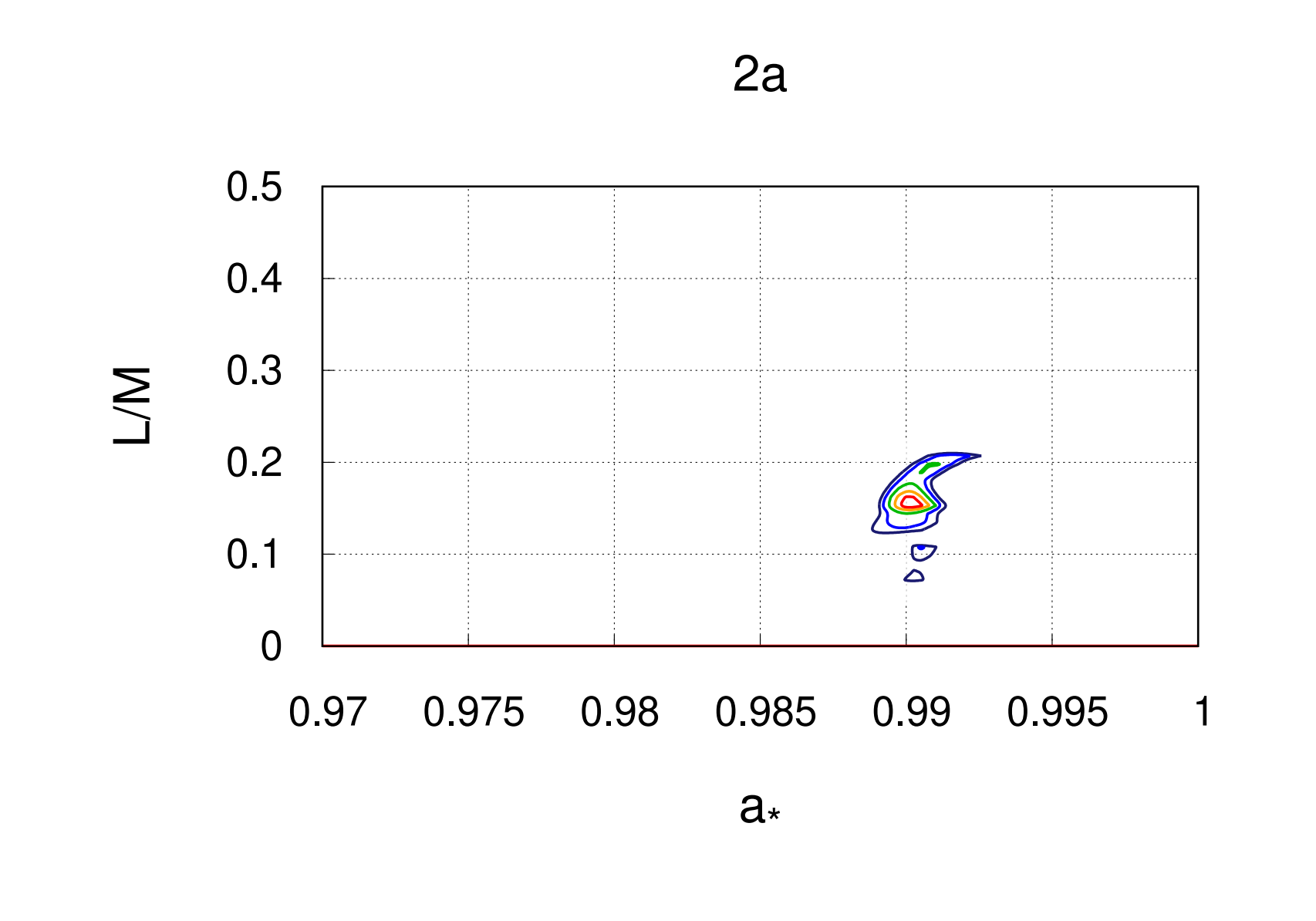} \hspace{0.5cm}
\includegraphics[width=0.48\textwidth,trim=2.5cm 2.0cm 2.5cm 0.0cm,clip]{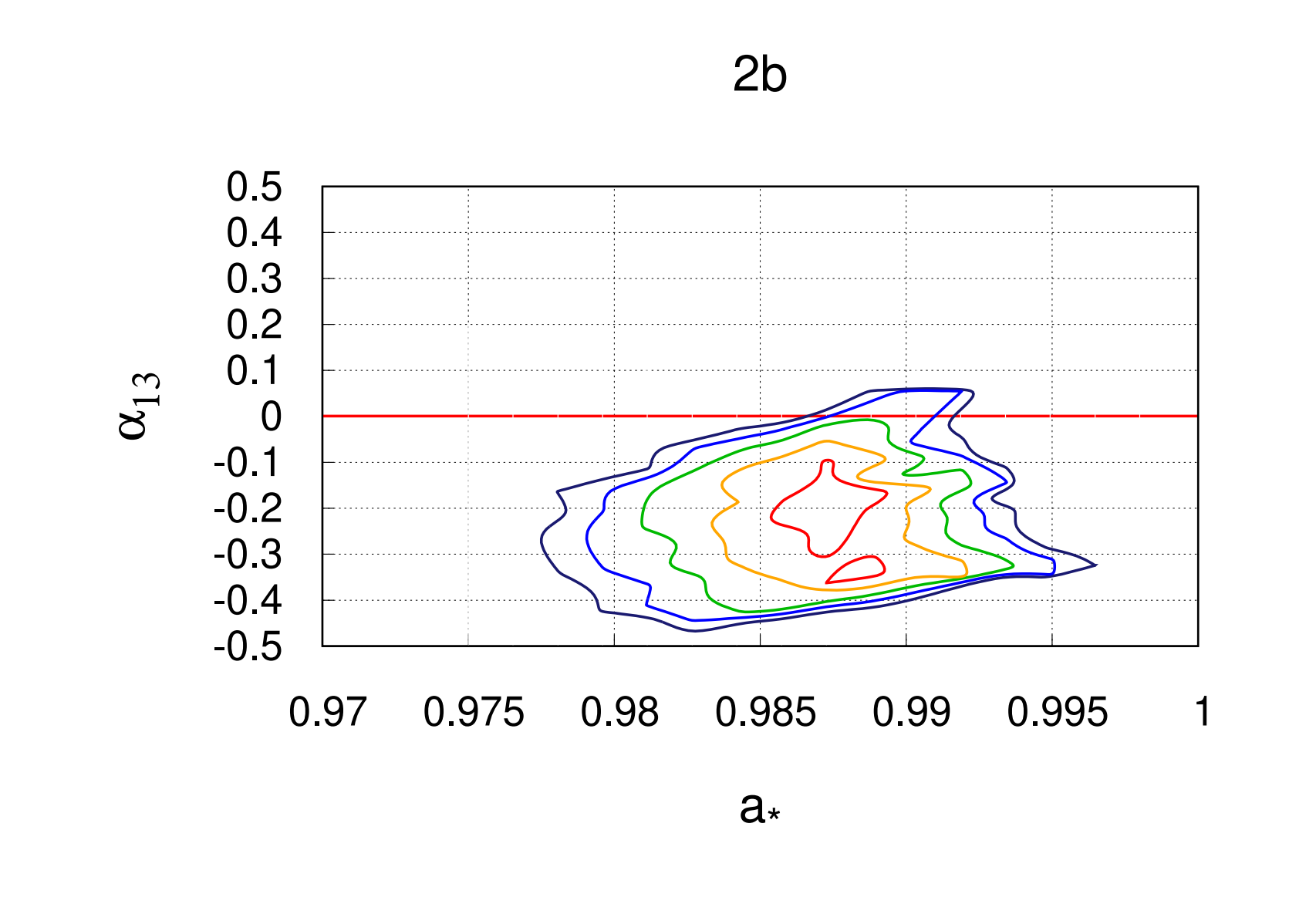}
\caption{Constraints on the black hole spin parameter $a_*$ and the deformation parameters $L/M$ and $\alpha_{13}$ from Fit~1a (top-left panel), Fit~1b (top-right panel), Fit~2a (bottom-left panel), and Fit~2b (bottom-right panel). The red, yellow, green, blue, and black curves represent, respectively, the 1-, 2-, 3-, 4-, and 5-$\sigma$ confidence level contours after marginalizing over all other free parameters. The horizontal red solid lines at $L/M =0$ and $\alpha_{13} = 0$ correspond to the Kerr solution of general relativity. See the text for more details.}
\label{f-12-c}
\end{figure*}

\begin{figure*}[t]
\centering
\includegraphics[width=0.48\textwidth,trim=0cm 0cm 0cm 0cm,clip]{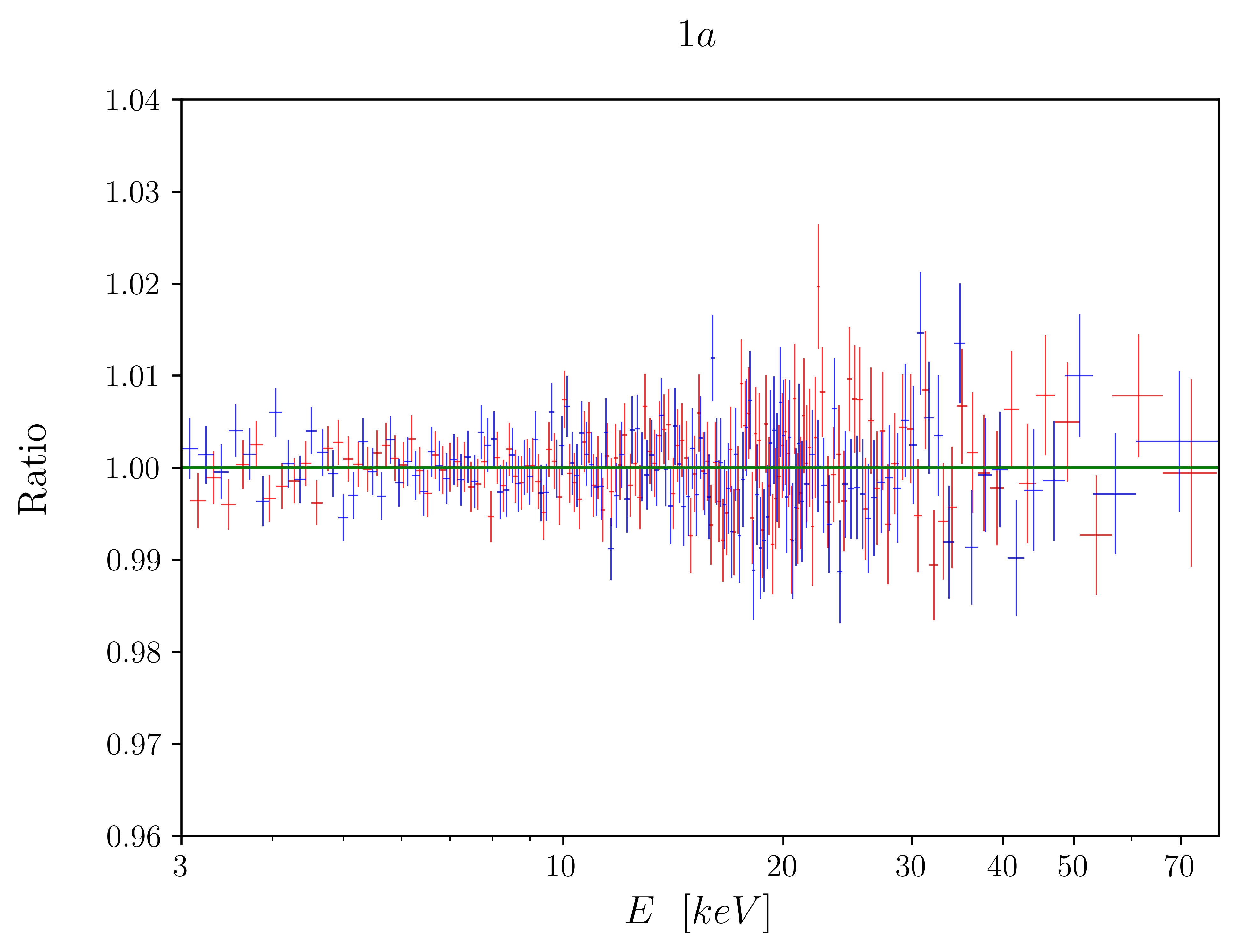} \hspace{0.5cm}
\includegraphics[width=0.48\textwidth,trim=0cm 0cm 0cm 0cm,clip]{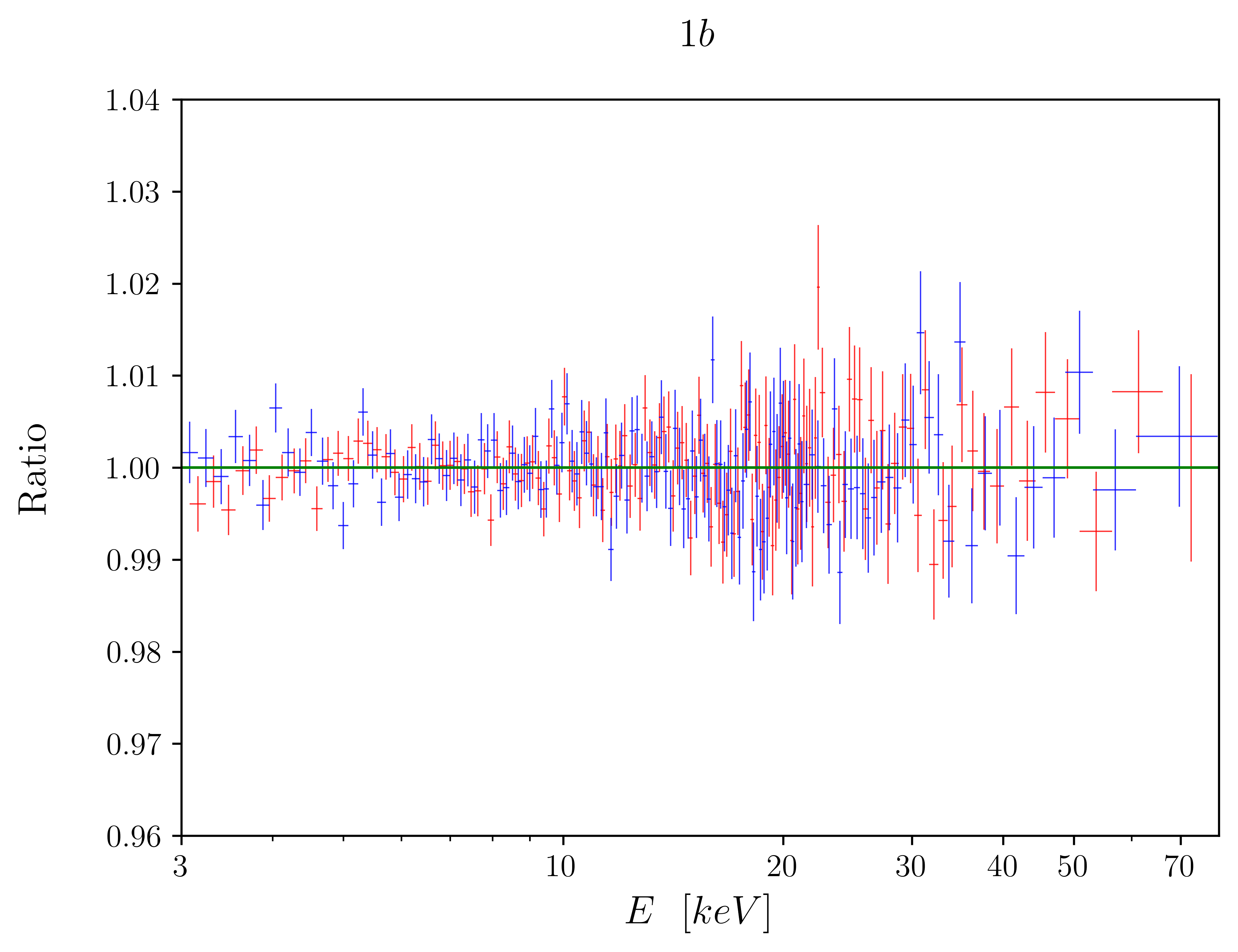}
\\
\vspace{0.5cm}
\includegraphics[width=0.48\textwidth,trim=0cm 0cm 0cm 0cm,clip]{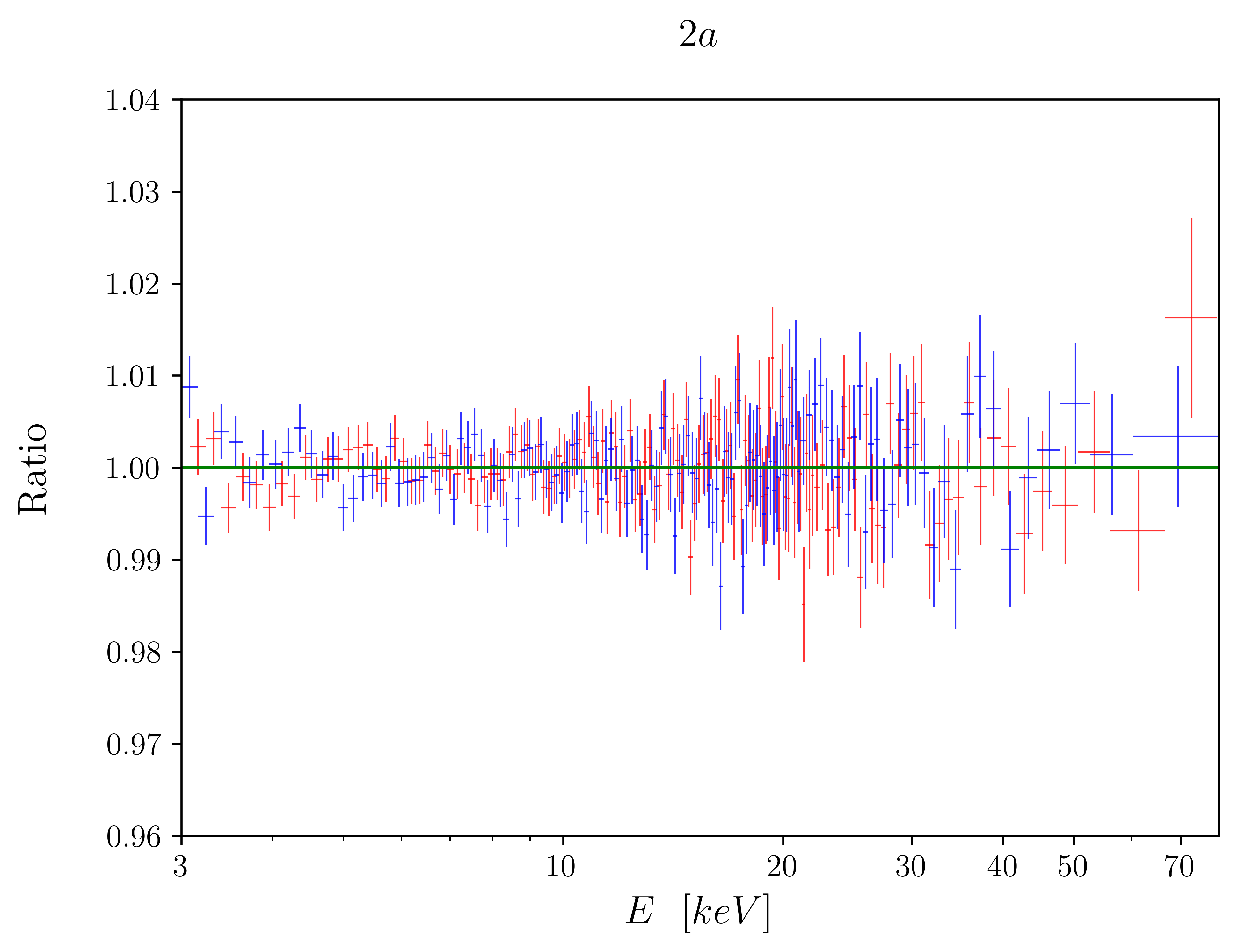} \hspace{0.5cm}
\includegraphics[width=0.48\textwidth,trim=0cm 0cm 0cm 0cm,clip]{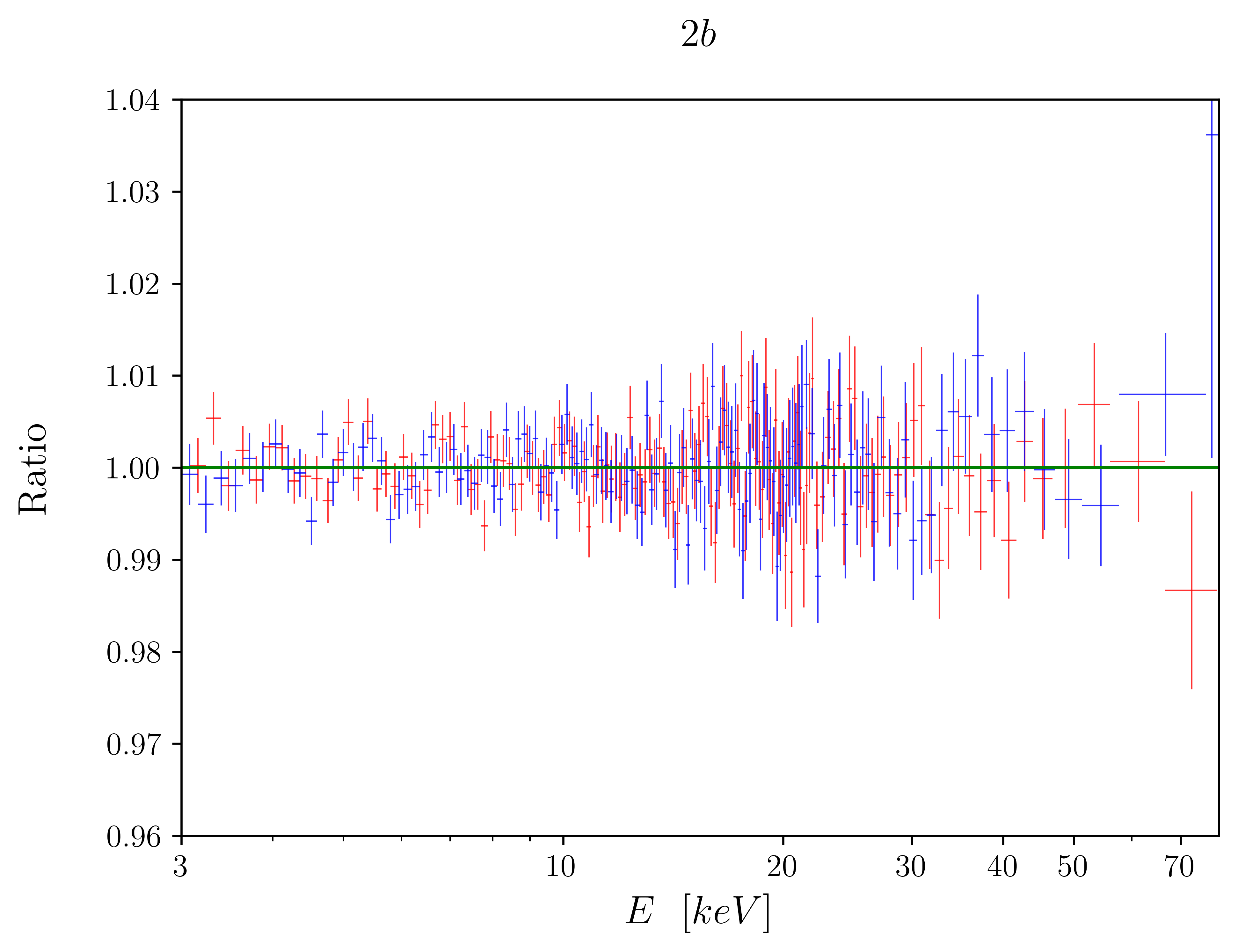}
\caption{Data to best-fit model ratios for Fit~1a (top-left panel), Fit~1b (top-right panel), Fit~2a (bottom-left panel), and Fit~2b (bottom-right panel). Red and blue data are, respectively, for FPMA and FPMB, which are the two detectors onboard \textsl{NuSTAR}. See the text for more details.}
\label{f-12-r}
\end{figure*}

\vspace{0.2cm}

\noindent where {\tt constant} is a cross-calibration constant to account for the difference between the two \textsl{NuSTAR} detectors, FPMA and FPMB. We set the constant for FPMA, $C_{\rm FPMA}$, to 1, and leave the constant for FPMB, $C_{\rm FPMB}$, free in the fit. We start by fitting the simulated spectrum with the version of {\tt relxill\_nk} used for the simulations and therefore correctly assuming that the spacetime metric is described by the black hole metric in conformal gravity. This is Fit~1a. In the fit, we leave free the following parameters: $a_*$, $L/M$, $\Gamma$, $i$, $q$, $\log\xi$, $A_{\rm Fe}$, $R_{\rm f}$, the normalization, and $C_{\rm FPMB}$. $E_{\rm cut}$ is frozen to the input value 300~keV because its impact is in any case weak, since \textsl{NuSTAR} covers the energy band 3 to 79~keV. The estimates of the model parameters from our fit are reported in Tab.~\ref{t-fit12} (second column). As expected we recover the correct input parameters and the fit is good. The top-left panel in Fig.~\ref{f-12-c} shows the 1-, 2-, 3-, 4-, and 5-$\sigma$ confidence level curves on the plane spin parameter vs deformation parameter after marginalizing over all other free parameters of the model. As we can see, from a similar analysis we could safely rule out the Kerr solution, which is at $L/M = 0$ and is marked by the horizontal red solid line in Fig.~\ref{f-12-c}. The top-left panel in Fig.~\ref{f-12-r} shows the residuals of this fit and we do not see particular features, namely the fit is good.

Let us now assume that we have observed such a black hole in conformal gravity with \textsl{NuSTAR}, but we do not know it is a black hole in conformal gravity and therefore we follow an agnostic strategy. This time we fit the simulated data with the version of {\tt relxill\_nk} employing the Johannsen metric with non-vanishing deformation parameter $\alpha_{13}$, whose line element is reported in Appendix~\ref{aaa-3}. So far this has been the most widely used metric for X-ray tests of the Kerr hypothesis. This is Fit~1b. As in the case of Fit~1a, we have 10~free parameters in the fit. The best fit values are reported in Tab.~\ref{t-fit12} (fourth column). The top-right panel in Fig.~\ref{f-12-c} shows the 1-, 2-, 3-, 4-, and 5-$\sigma$ confidence level curves on the plane spin parameter vs deformation parameter after marginalizing over all other free parameters. Now the Kerr solution corresponds to $\alpha_{13} = 0$ and, as we can see from the plot, we can rule out the Kerr metric at more than 5-$\sigma$. The residuals of this fit are shown in the top-right panel in Fig.~\ref{f-12-r}: even in this case the fit looks good.

\begin{table*}
\centering
\renewcommand\arraystretch{1.5}
\begin{tabular}{lcccc}
\hline\hline
Model & 1a & 2a & 1b & 2b\\
\hline

$a_*$ & $0.9908_{-0.0008}^{+0.0007}$ & $0.9902_{-0.0004}^{+0.0004}$ & $0.9872_{-0.005}^{+0.0026}$ & $0.9877_{-0.002}^{+0.0018}$ \\

$L/M$ & $0.212_{-0.005}^{+0.002}$ & $0.1536_{-0.0004}^{+0.0070}$ & -- &  -- \\

$\alpha_{13}$ & -- & -- & $-0.22_{-0.08}^{+0.07}$ &  $-0.16_{-0.04}^{+0.03}$ \\

$\Gamma$ & $1.700_{-0.002}^{+0.003}$ & $1.704_{-0.002}^{+0.001}$ & $1.702_{-0.002}^{+0.001}$ & $1.700_{-0.005}^{+0.003}$ \\

$i$ [deg] & $62.6_{-0.9}^{+0.9}$ & $59.9_{-0.1}^{+0.3}$ & $63.5_{-0.3}^{+2.5}$ & $61.4_{-0.5}^{+1.2}$ \\

$q$ & $8.81_{-0.07}^{+0.09}$ & $7.97_{-0.09}^{+0.05}$ & $8.32_{-0.24}^{+0.50}$ & $8.01_{-0.06}^{+0.40}$ \\

$\log\xi$ [erg~cm~s$^{-1}$] & $1.98_{-0.07}^{+0.07}$ & $1.95_{-0.04}^{+0.02}$ & $2.00_{-0.03}^{+0.04}$ & $2.02_{-0.01}^{+0.05}$ \\

A$_{\rm Fe}$ & $0.996_{-0.009}^{+0.030}$ & $1.01_{-0.02}^{+0.08}$  & $1.00_{-0.01}^{+0.04}$ & $1.046_{-0.006}^{+0.070}$ \\

$R_{\rm f}$ & $1.98_{-0.06}^{+0.06}$ & $1.99_{-0.02}^{+0.08}$ & $2.01_{-0.05}^{+0.05}$ & $2.0315_{-0.0500}^{+0.0001}$ \\

Norm & $0.0337_{-0.0005}^{+0.0004}$ & $0.0332_{-0.0004}^{+0.0002}$ & $0.0336_{-0.0001}^{+0.0004}$ & $0.0327_{-0.0010}^{+0.0004}$ \\

$C_{\rm FPMB}$ & $0.9993_{-0.0007}^{+0.0007}$ &  $0.9997_{-0.0007}^{+0.0007}$ & $0.9993_{-0.0007}^{+0.0007}$ & $1.0005_{-0.0007}^{+0.0007}$\\

\hline

$\chi^2$/dof & 541.05/580 & 515.48/578 & 549.71/580 & 565.2/578 \\

& = 0.93284 & = 0.89183 & = 0.94778 & = 0.97785 \\

\hline\hline
\end{tabular}
\caption{Summary of the best-fit values for models 1a, 2a, 1b, and 2b. The reported uncertainties correspond to 90\% confidence level for one relevant parameter ($\Delta\chi^2 = 2.71$).} 
\label{t-fit12}
\end{table*}

We proceed as in the case of Simulations~1 and we analyze Simulation~2, which is characterized by a lower value of $L/M$; i.e., it is closer to the Kerr solution. First, we fit the simulated data with the version of {\tt relxill\_nk} for black holes in conformal gravity. This is Fit~2a. The estimates of the model parameters are reported in Tab.~\ref{t-fit12} (third column) and the bottom-left panel in Fig.~\ref{f-12-c} shows the 1-, 2-, 3-, 4-, and 5-$\sigma$ confidence level curves on the plane spin parameter vs deformation parameter after marginalizing over all other free parameters of the model. The bottom-left panel in Fig.~\ref{f-12-r} shows the residuals of this fit. There are no surprises: the fit is good and we can still rule out the Kerr solution at more than 5-$\sigma$.

Last, we fit Simulation~2 with the version of {\tt relxill\_nk} employing the Johannsen metric with non-vanishing deformation parameter $\alpha_{13}$. This is Fit~2b. The results are shown in the last column in Tab.~\ref{t-fit12} (last column) and in the bottom-right panels in, respectively, Fig.~\ref{f-12-c} and Fig.~\ref{f-12-r}. While the fit is still good as we do not see significant features in the residual plot, we cannot rule out the Kerr solution at 4- or 5-$\sigma$ any longer. We can still rule out the Kerr solution at 3-$\sigma$.


\subsection{Einstein-Maxwell-dilaton-axion gravity}\label{ss-2}

\begin{figure*}[t]
\centering
\includegraphics[width=0.48\textwidth,trim=2.5cm 1.0cm 2.5cm 0.0cm,clip]{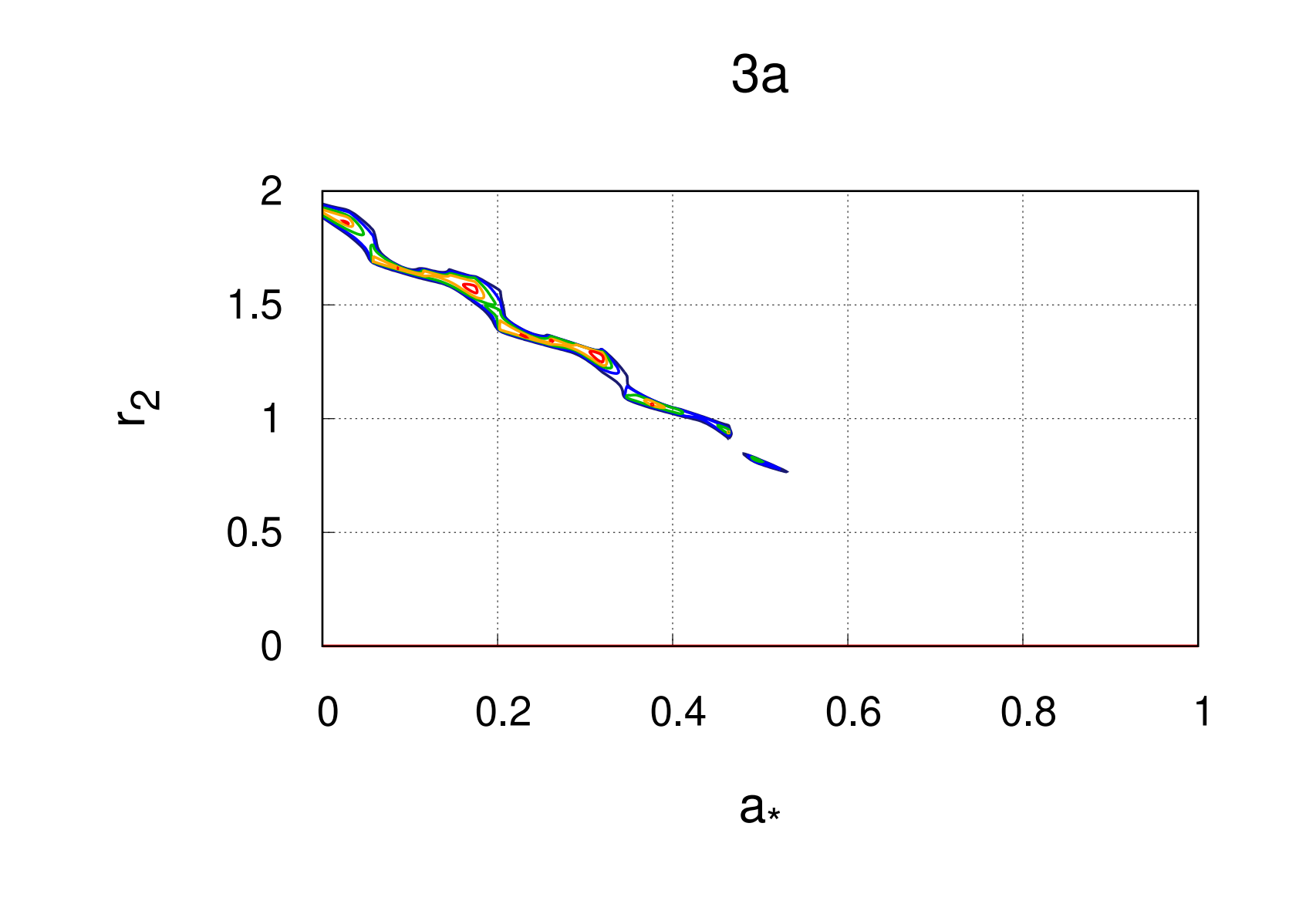} \hspace{0.5cm}
\includegraphics[width=0.48\textwidth,trim=2.5cm 1.0cm 2.5cm 0.0cm,clip]{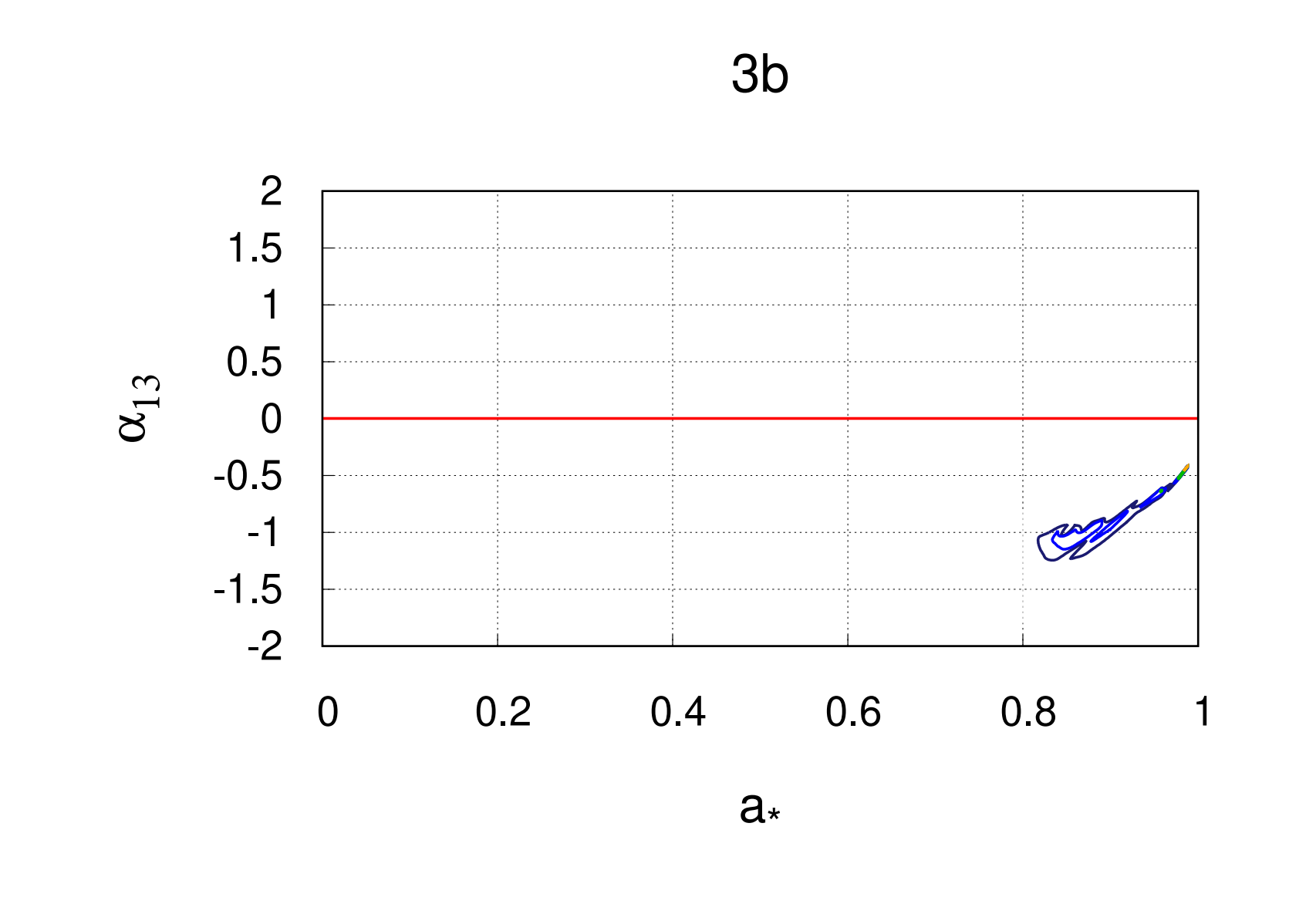} 
\includegraphics[width=0.48\textwidth,trim=2.5cm 2.0cm 2.5cm 0.0cm,clip]{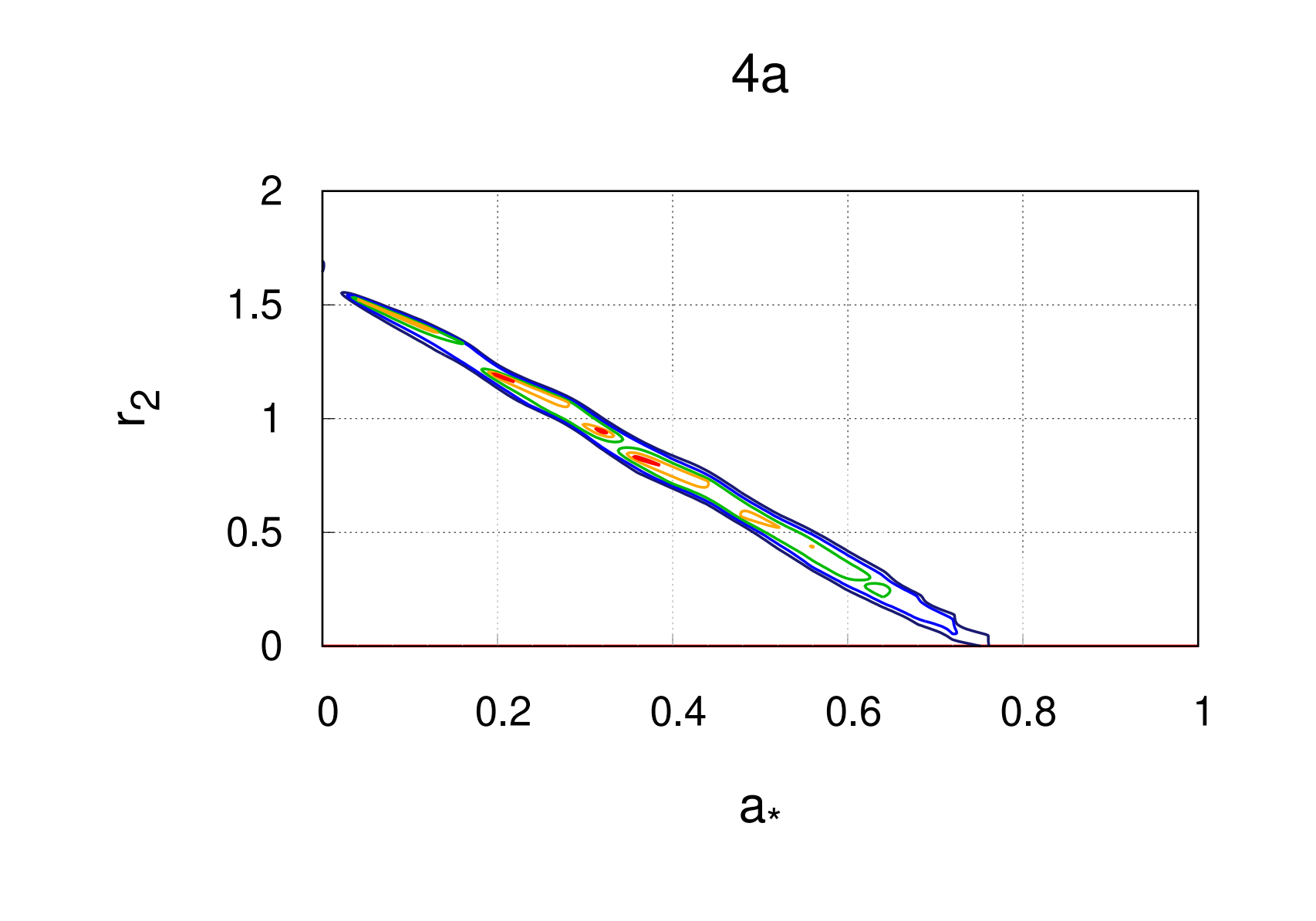} \hspace{0.5cm}
\includegraphics[width=0.48\textwidth,trim=2.5cm 2.0cm 2.5cm 0.0cm,clip]{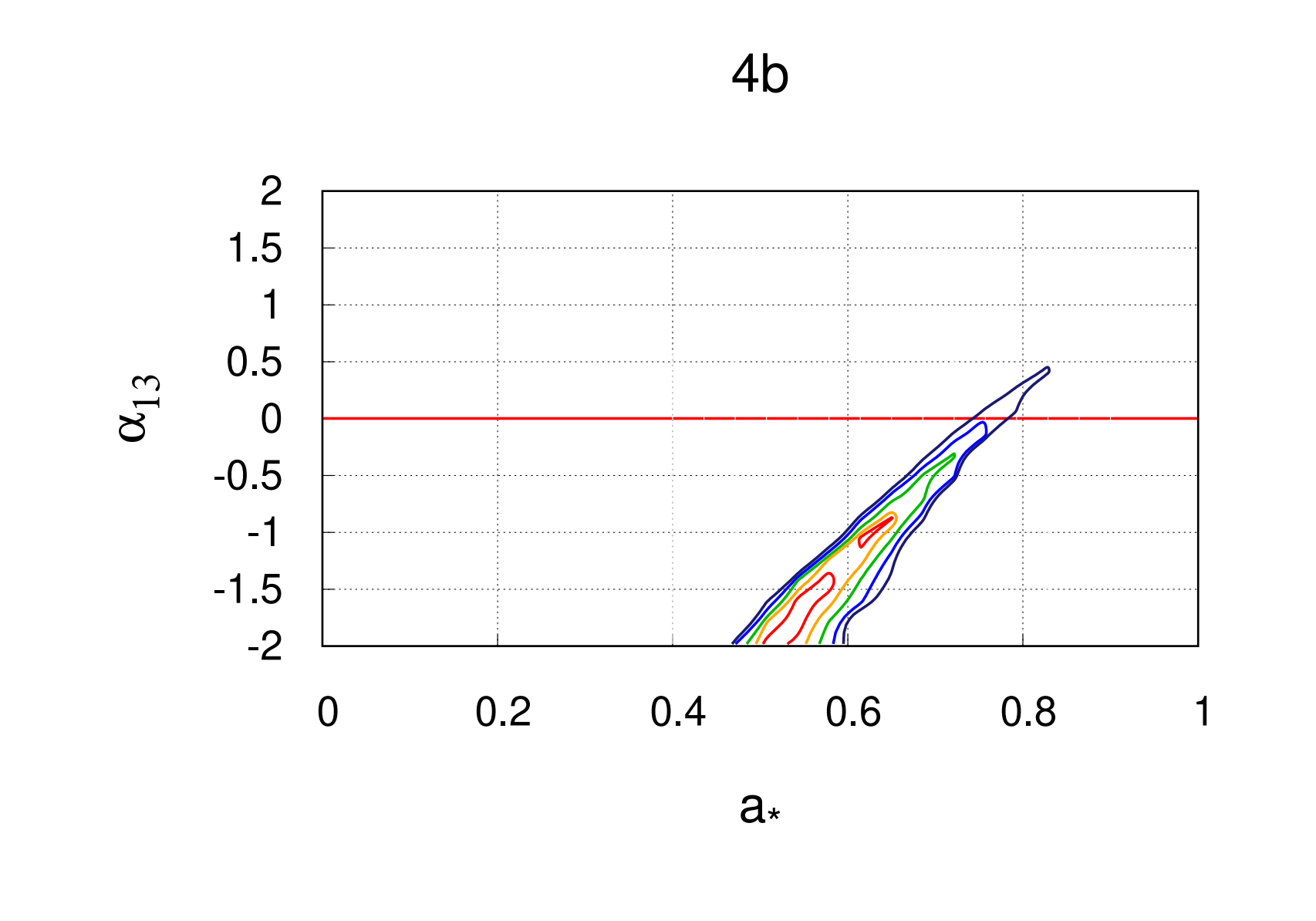}
\caption{Constraints on the black hole spin parameter $a_*$ and the deformation parameters $r_2$ and $\alpha_{13}$ from Fit~3a (top-left panel), Fit~3b (top-right panel), Fit~4a (bottom-left panel), and Fit~4b (bottom-right panel). The red, yellow, green, blue, and black curves represent, respectively, the 1-, 2-, 3-, 4-, and 5-$\sigma$ confidence level contours after marginalizing over all other free parameters. The horizontal red solid lines at $r_2 =0$ and $\alpha_{13} = 0$ correspond to the Kerr solution of general relativity. See the text for more details.}
\label{f-34-c}
\end{figure*}

\begin{figure*}[t]
\centering
\includegraphics[width=0.48\textwidth,trim=0cm 0cm 0cm 0cm,clip]{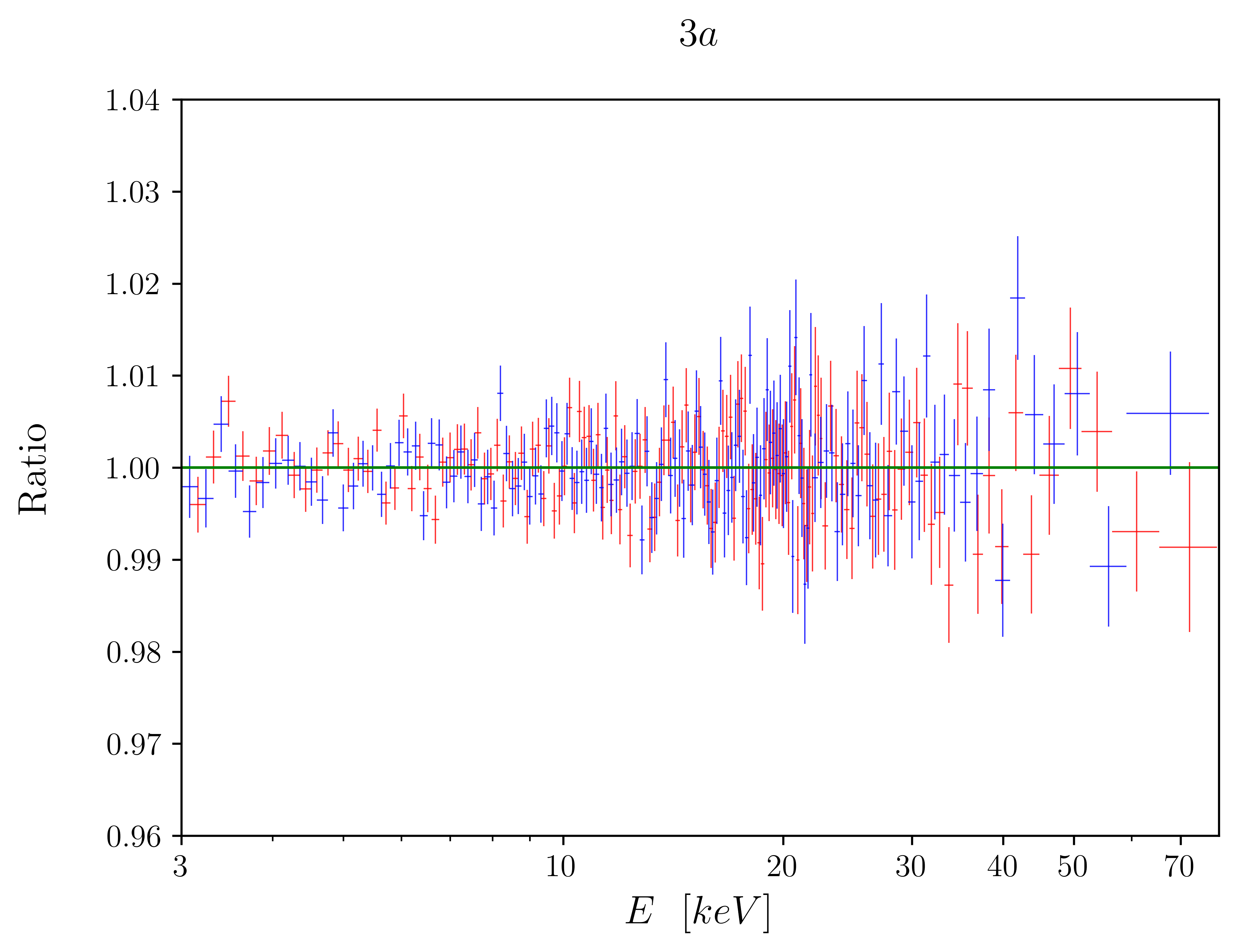} \hspace{0.5cm}
\includegraphics[width=0.48\textwidth,trim=0cm 0cm 0cm 0cm,clip]{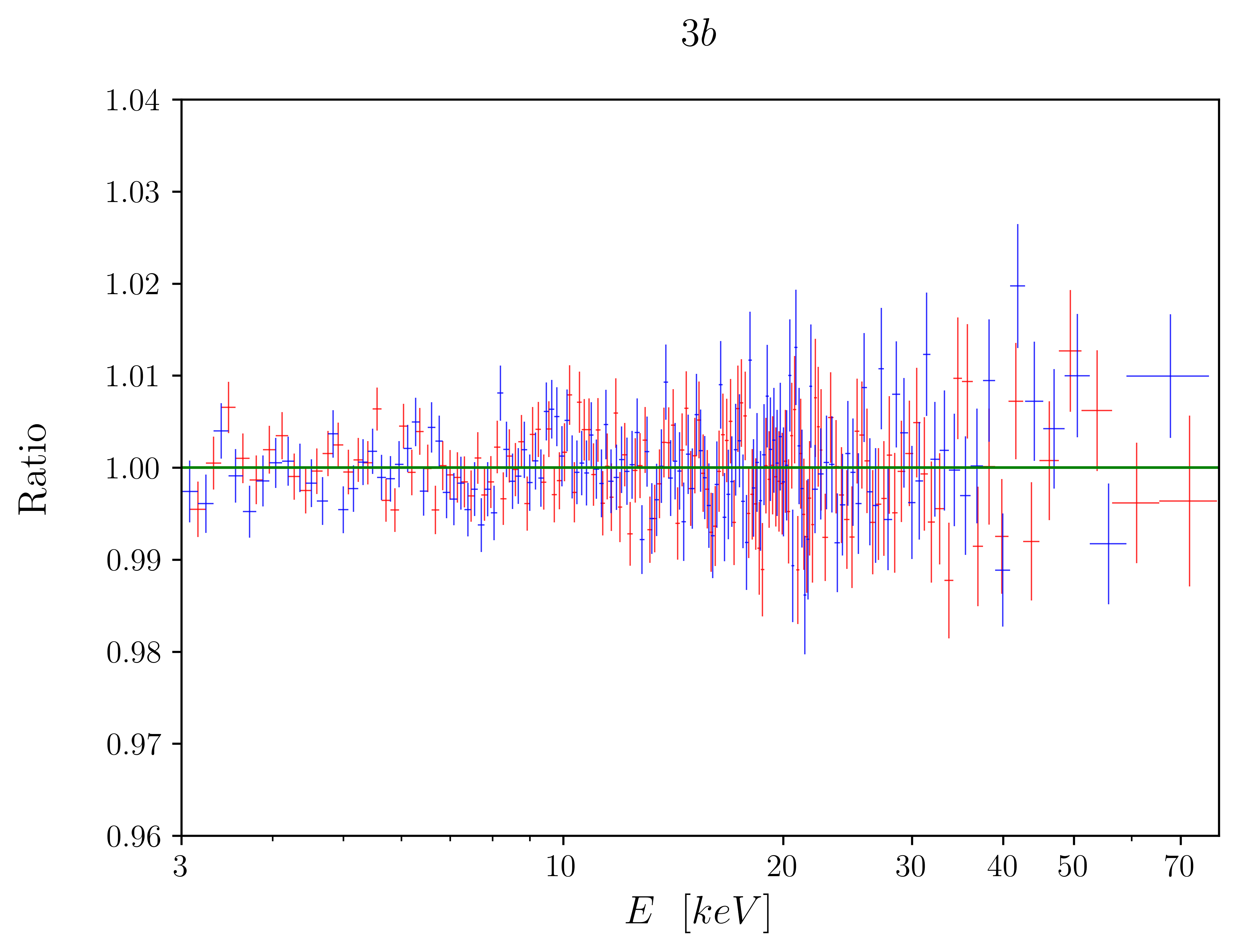}
\\
\vspace{0.5cm}
\includegraphics[width=0.48\textwidth,trim=0cm 0cm 0cm 0cm,clip]{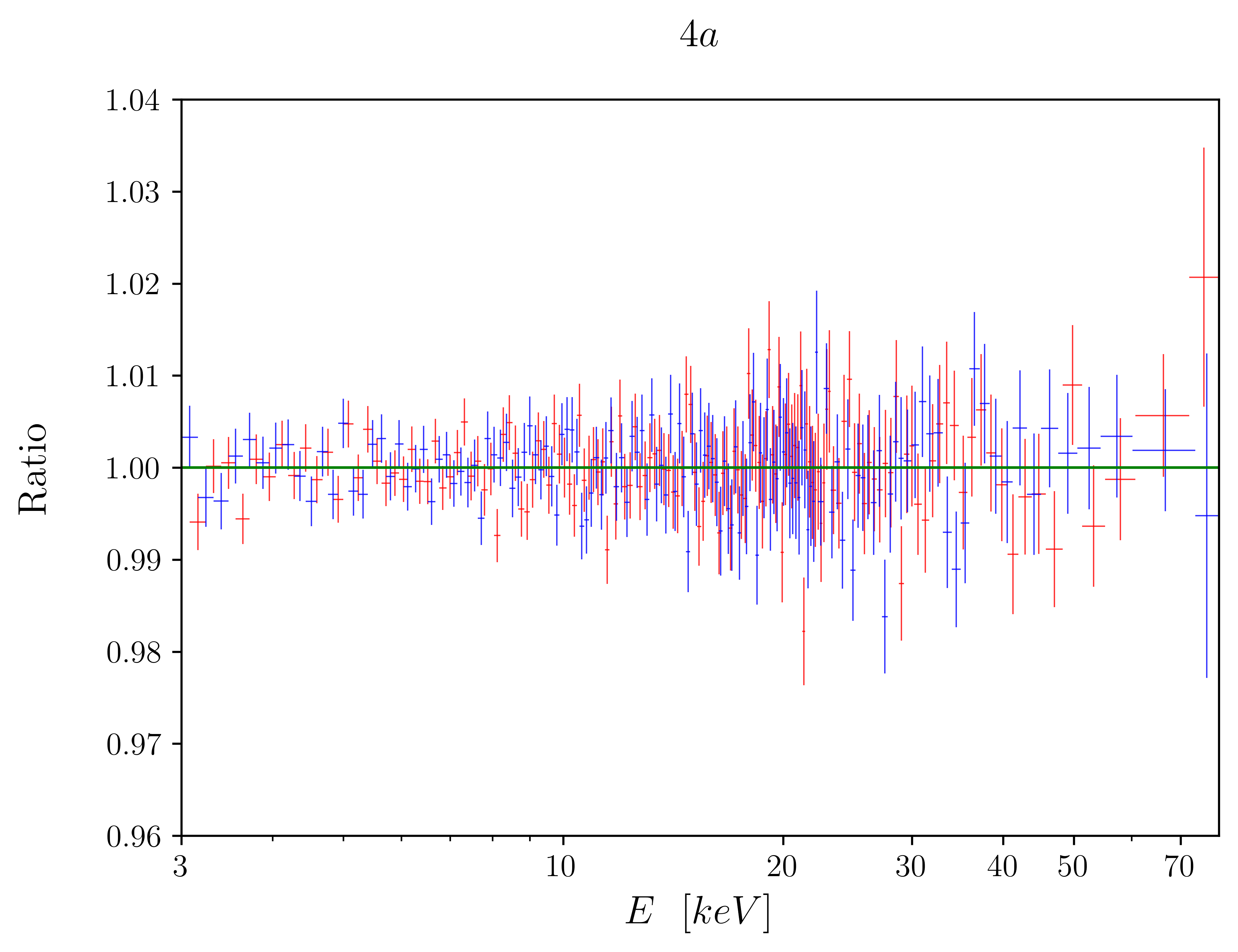} \hspace{0.5cm}
\includegraphics[width=0.48\textwidth,trim=0cm 0cm 0cm 0cm,clip]{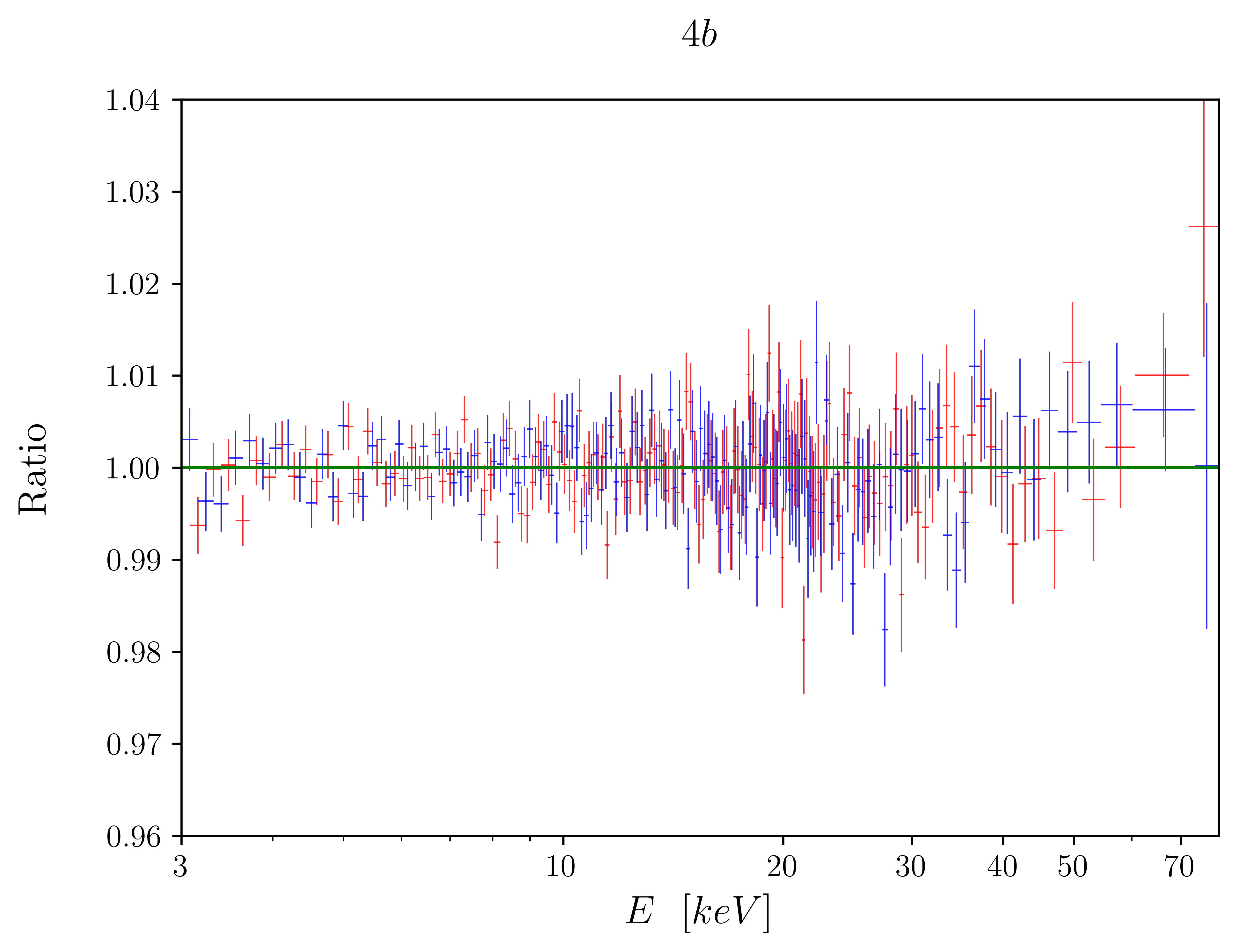}
\caption{Data to best-fit model ratios for Fit~3a (top-left panel), Fit~3b (top-right panel), Fit~4a (bottom-left panel), and Fit~4b (bottom-right panel). Red and blue data are, respectively, for FPMA and FPMB, which are the two detectors onboard \textsl{NuSTAR}. See the text for more details.}
\label{f-34-r}
\end{figure*}

As our second case study, we consider black holes in Einstein-Maxwell-dilaton-axion gravity. The metric is reported in Appendix~\ref{aaa-2}. The spacetime has three parameters: the mass of the object, $\mathcal{M}$, the dimensionless spin parameter, $a_*$, and the dimensionless dilaton parameter, $r_2$. The Kerr solution is recovered when the dilaton parameter vanishes, $r_2 = 0$. The allowed range of $r_2$ is
\be
0 \le r_2 \le 2 \left( 1 - \left| a_* \right| \right) \, .
\ee
The implementation of this black hole spacetime in {\tt relxill\_nk} was presented in Ref.~\cite{Tripathi:2021rwb}.

As in the case of the simulations in conformal gravity, we want to consider sources in which the relativistic effects are strong, which, in turn, requires that the inner edge of the accretion disk is very close to the black hole event horizon. As shown in Fig.~1 of Ref.~\cite{Tripathi:2021rwb}, we should choose black holes with the value of $r_2$ close to the its maximum value $r_2^{\rm max} = 2 \left( 1 - \left| a_* \right| \right)$. Deviations from general relativity turn out to be larger for low values of the spin parameter $a_*$ and high values of the dilaton parameter $r_2$. For our simulations, we choose $a_* = 0.2$ and $r_2 = 1.4$ (Simulation~3) and 1.2 (Simulation~4).

We proceed as in the spectral analysis of the black holes in conformal gravity. First, we fit the spectrum of Simulation~3 with the correct model, so we use the version of {\tt relxill\_nk} in which the spacetime metric is described by the black hole solution in Einstein-Maxwell-dilaton-axion gravity. This is Fit~3a. The results of the fit are reported in Tab.~\ref{t-fit34} (second column). The top-left panel in Fig.~\ref{f-34-c} shows the 1-, 2-, 3-, 4-, and 5-$\sigma$ confidence level curves on the plane spin parameter vs deformation parameter after marginalizing over all other free parameters of the model. The Kerr solution is ruled out at more than 5-$\sigma$. The residuals of the fit are shown in the top-left panel in Fig.~\ref{f-34-r}.

We fit the spectrum of Simulation~3 with the version of {\tt relxill\_nk} employing the Johannsen spacetime with deformation parameter $\alpha_{13}$ (Fit~3b). The estimates of the model parameters are reported in Tab.~\ref{t-fit34} (fourth column). The constraints on $a_*$ and $\alpha_{13}$ are shown in the top-right panel in Fig.~\ref{f-34-c} and the residuals of the fit are reported in the top-right panel in Fig.~\ref{f-34-r}. The quality of the fit is good and we can rule out the Kerr metric at more than 5-$\sigma$.

We move to the spectral analysis of Simulation~4 and we fit the spectrum with the correct model (Fit~4a). The estimates of the model parameters are reported in Tab.~\ref{t-fit34} (third column). The constraints on $a_*$ and $r_2$ are shown in the bottom-left panel in Fig.~\ref{f-34-c} and the residuals of the fit are reported in the bottom-left panel in Fig.~\ref{f-34-r}. Now we can rule out the Kerr solution at 4-$\sigma$ but not at 5-$\sigma$ even if we use the correct background metric. As discussed in Ref.~\cite{Tripathi:2021rwb}, it is very challenging to distinguish black holes in Einstein-Maxwell-dilaton-axion gravity from the Kerr black holes in general relativity and eventually we can do it only for near extremal objects.

Last, we fit the spectrum of Simulation~4 with the version of {\tt relxill\_nk} employing the Johannsen spacetime with deformation parameter $\alpha_{13}$. This is Fit~4b. The best-fit values are reported in Tab.~\ref{t-fit34} (last column). The constraints on $a_*$ and $\alpha_{13}$ are shown in the bottom-right panel in Fig.~\ref{f-34-c} and the residuals of the fit are reported in the bottom-right panel in Fig.~\ref{f-34-r}. As in Fit~4a, we can rule out the Kerr solution at 4-$\sigma$ but not at 5-$\sigma$. The quality of the fit is good and we do not see large residuals in the bottom-right panel in Fig.~\ref{f-34-r}.

\begin{table*}
\centering
\renewcommand\arraystretch{1.5}
\begin{tabular}{lcccc}
\hline\hline
Model & 3a & 4a & 3b & 4b\\
\hline

$a_{*}$ & $0.329_{-0.013}^{+0.014}$ & $0.23_{-0.13}^{+0.29}$ & $0.987_{-0.001}^{+0.002}$ & $0.56_{-0.07}^{+0.10}$\\

$r_2$ & $1.22_{-0.14}^{+0.07}$ & $1.12_{-0.59}^{+0.27}$ & - &  - \\

$\alpha_{13}$ & - & - & $-0.42_{-0.02}^{+0.02}$ & $-1.52_{-0.63}^{+1.52}$ \\

$\Gamma$ & $1.701_{-0.002}^{+0.003}$ & $1.701_{-0.001}^{+0.001}$ & $1.717_{-0.002}^{+0.002}$ & $1.705_{-0.003}^{+0.003}$\\

$i$ [deg] & $59.9_{-1.7}^{+2.5}$ & $59.4_{-0.9}^{+0.9}$ & $65.2_{-0.1}^{+0.8}$ & $58.2_{-0.8}^{+1.2}$\\

$q$ & $9.2_{-0.7}$ & $7.7_{-0.4}^{+0.5}$ & $10.0_{-0.1}$ & $8.5_{-0.7}^{+0.8}$\\

$\log\xi$ [erg~cm~s$^{-1}$] & $1.98_{-0.06}^{+0.04}$ & $2.000_{-0.060}^{+0.008}$ & $1.91_{-0.04}^{+0.04}$ & $1.95_{-0.04}^{+0.05}$\\

A$_{\rm Fe}$ & $0.968_{-0.024}^{+0.012}$ & $1.014_{-0.018}^{+0.012}$ & $0.864_{-0.004}^{+0.004}$ & $0.978_{-0.012}^{+0.011}$\\

$R_{\rm f}$ & $1.98_{-0.04}^{+0.04}$ & $2.00_{-0.02}^{+0.01}$ & $1.95_{-0.03}^{+0.02}$ & $1.97_{-0.02}^{+0.04}$\\

Norm & $0.0346_{-0.0003}^{+0.0003}$ & $0.0355_{-0.0002}^{+0.0002}$ & $0.0332_{-0.0002}^{+0.0003}$ & $0.0353_{-0.0003}^{+0.0002}$\\

$C_{\rm FPMB}$ & $0.9992_{-0.0007}^{+0.0007}$ & $0.9994_{-0.0007}^{+0.0007}$ & $0.9992_{-0.0007}^{+0.0007}$ & $0.9994_{-0.0007}^{+0.0007}$\\

\hline

$\chi^2$/dof & 610.31/582 & 552.61/589 & 633.37/582 & 562.58/589 \\

& = 1.04864 & = 0.93822 & = 1.08826 & = 0.95514 \\

\hline\hline
\end{tabular}
\caption{Summary of the best-fit values for models 3a, 4a, 3b, and 4b. The reported uncertainties correspond to 90\% confidence level for one relevant parameter ($\Delta\chi^2 = 2.71$).} 
\label{t-fit34}
\end{table*}


\section{Discussion and conclusions}\label{s-d}

Most of the tests of the Kerr hypothesis reported in the literature, and in particular most of the X-ray tests of the Kerr hypothesis, follow an agnostic approach in which one considers some parametric black hole spacetime (which is not a solution of any gravity theory and is simply obtained by deforming the Kerr metric under certain requirements) and tries to constrain the value(s) of its deformation parameter(s) by comparing theoretical predictions with observations. The spirit of this strategy is to perform a null experiments in which we expect to recover the predictions of general relativity but we can potentially measure even deviations from Einstein's gravity. The question is whether such an approach is indeed suitable to discover new physics.

We cannot have a universal answer to this question because it certainly depends on the particular spacetime metric around the source, on our choice of the parametric black hole spacetime and of the deformation parameter for our data analysis, and, last but not least, on the specific values of the parameters of the source (like inclination angle of the disk, metallicity of the disk, ionization of the disk, etc. which have nothing to do with gravity). In this work, we have presented the results for two case studies, black holes in conformal gravity and black holes in Einstein-Maxwell-dilaton-axion gravity, and we have used the Johannsen spacetime for the agnostic tests. In all our simulations, we have considered systems in which the inner edge of the accretion disk is very close to the black hole and the corona illuminates mainly the very inner part of the disk. These are the requirements to get accurate and precise measurements from the analysis of relativistically blurred reflection features, not only in the case of tests of the Kerr hypothesis but even for spin measurements that assume general relativity.

In our work, we have not considered the systematic uncertainties related to the astrophysical model. There are a number of studies published in the literature supporting the conclusion that, if we select properly the sources and the observations to analyze, the analysis of relativistic reflection features in the available X-ray data with current reflection models can provide precise and accurate measurements (see, for instance, the discussion in~\cite{Bambi:2022dtw} and references therein). The uncertainties in current measurements are dominated by statistical errors, while systematic errors due to the reflection model are subdominant. With this spirit, our work has only investigated if an agnostic approach has the ability to discover new physics. If we do not employ the correct astrophysical model or we do not select properly the sources and the observations to analyze, the final measurements can be strongly biased even if we use the correct background metric~\cite{Riaz:2019kat}.

From the analysis of the spectra of black holes in conformal gravity, we can argue that agnostic tests can still discover new physics but they are likely somewhat less powerful than theory-specific tests employing the correct background metric. However, it is worth pointing out that there are many theories of gravity and it would be certainly unlikely to use the correct one for our tests. We also note that we can get very good fits even with an incorrect metric (as we can see from the residual plots), so an evaluation of the quality of the fit cannot help to discover new physics.

On the other hand, from the analysis of the spectra of black holes in Einstein-Maxwell-dilaton-axion gravity, we have not found a clear difference in the constraining power of the theory-specific and agnostic tests. Even in this case, the quality of the best-fits is always good and therefore it does not give any hint on the fact we are using the correct metric or an incorrect one.

In our simulations, we have assumed the observation of a bright Galactic black hole with \textsl{NuSTAR}. The next generation of X-ray missions (\textsl{eXTP}~\cite{eXTP:2016rzs}, \textsl{Athena}~\cite{Nandra:2013jka}, \textsl{HEX-P}, etc.) promise to provide unprecedented high-quality data, which can potentially permit more precise tests of general relativity. With such high-quality data, it should be easier to break the parameter degeneracy and therefore the gap of the constraining power between theory-specific and agnostic tests may be reduced. On the other hand, they will introduce new challenges because they will necessary require more accurate synthetic spectra than those available today and it may be more difficult to distinguish a different astrophysical environment from a deviation from general relativity without the development of sufficiently advanced models.


\vspace{0.5cm}

{\bf Acknowledgments --}
This work was supported by the National Natural Science Foundation of China (NSFC), Grant No.~12250610185, 11973019, and 12261131497, the Natural Science Foundation of Shanghai, Grant No. 22ZR1403400, and the Shanghai Municipal Education Commission, Grant No. 2019-01-07-00-07-E00035. S.R. acknowledges also the support from the China Postdoctoral Science Foundation, Grant No. 2022M720035, and the Teach@T\"ubingen Fellowship.


\appendix

\section{Black hole metrics}\label{aaa}

For convenience, we report below the black hole metrics used in this work. More details can be found in the original papers. We employ natural units in which $c = G_{\rm N} = 1$ and metrics with signatures $(-+++)$.

\subsection{Black holes in conformal gravity}\label{aaa-1}

In Boyer-Lindquist coordinates, the line element of the black holes in conformal gravity discussed in Ref.~\cite{Bambi:2016wdn} is
\be
ds^2 = \left( 1 + \frac{L^2}{\Sigma} \right)^2 ds^2_{\rm Kerr} \, , \nonumber
\ee
where $L$ is the conformal parameter and $ds^2_{\rm Kerr}$ is the standard line element of the Kerr metric in Boyer-Lindquist coordinates. $L$ is non-negative and for $L=0$ we recover the Kerr solution. In our manuscript, we use the quantity $L/M$ because it is dimensionless.

\subsection{Black holes in Einstein-Maxwell-dilaton-axion gravity}\label{aaa-2}

In Boyer-Lindquist coordinates, the line element of the black holes in Einstein-Maxwell-dilaton-axion gravity is~\cite{Tripathi:2021rwb}
\be
ds^2 &=& - \left( 1 - \frac{2 \mathcal{M} r}{\tilde{\Sigma}} \right) dt^2 
+ \frac{\tilde{\Sigma}}{\Delta} \left( dr^2 + \Delta d\theta^2 \right) \nonumber\\
&& - \frac{4 a \mathcal{M} r}{\tilde{\Sigma}} \sin^2\theta \, dt d\phi \nonumber\\
&& + \sin^2\theta \left[ r \left( r + \tilde{r}_2 \right) + a^2 + \frac{2 \mathcal{M} r a^2 \sin^2\theta}{\tilde{\Sigma}} \right] d\phi^2 \, , \nonumber
\ee
where
\be
\tilde{\Sigma} &=& r \left( r + \tilde{r}_2 \right) + a^2 \cos^2\theta \, , \nonumber\\
\Delta &=& r \left( r + \tilde{r}_2 \right) - 2 \mathcal{M} r + a^2 \, , \nonumber
\ee
$\mathcal{M}$ is the mass of the object, $a$ is its specific spin angular momentum, and $\tilde{r}_2$ is the dilaton parameter. The condition for the existence of an event horizon is
\be
\left( \mathcal{M} - \frac{\tilde{r}_2}{2} \right)^2 - a^2 \ge 0 \, . \nonumber
\ee
The dimensionless spin parameter is $a_* = a/\mathcal{M}$ and the dimensionless dilaton parameter is $r_2 = \tilde{r}_2/\mathcal{M}$.

\subsection{Johannsen metric}\label{aaa-3}

In Boyer-Lindquist coordinates, the line element of Johannsen spacetime is~\cite{Johannsen:2013szh}
\be
ds^2 &=& - \frac{\tilde{\Sigma} \left( \Delta - a^2 A_2^2 \sin^2\theta\right)}{B^2} dt^2 \nonumber\\
&&+ \frac{\tilde{\Sigma}}{\Delta A_5} dr^2  + \tilde{\Sigma} d\theta^2 \nonumber\\
&& - \frac{2 a \left[ \left( r^2 + a^2\right) A_1 A_2 - \Delta \right] \tilde{\Sigma} \sin^2\theta}{B^2} dt d\phi \nonumber\\
&& + \frac{\left[\left( r^2 + a^2\right)^2 A_1^2 - a^2 \Delta \sin^2\theta \right] \tilde{\Sigma} \sin^2\theta}{B^2} d\phi^2 \, ,
\nonumber
\ee
where
\be
\tilde{\Sigma} &=& r^2 + a^2 \cos^2\theta + f \, , \nonumber\\
\Delta &=& r^2 - 2 M r + a^2 \, , \nonumber\\
B &=& \left( r^2 + a^2\right) A_1 - a^2 A_2 \sin^2\theta \, .
\ee
The functions $f$, $A_1$, $A_2$, and $A_5$ are defined as
\be
f &=& \sum_{n=3}^{\infty} \epsilon_n \frac{M^n}{r^{n-2}} \, , \nonumber\\
A_1 &=& 1 + \sum_{n=3}^{\infty} \alpha_{1n} \frac{M^n}{r^n} \, , \nonumber\\
A_2 &=& 1 + \sum_{n=2}^{\infty} \alpha_{2n} \frac{M^n}{r^n} \, , \nonumber\\
A_5 &=& 1 + \sum_{n=2}^{\infty} \alpha_{5n} \frac{M^n}{r^n} \, , \nonumber
\ee
The spacetime has four infinite sets of deformation parameters, $\{ \epsilon_n \}$, $\{ \alpha_{1n} \}$, $\{ \alpha_{2n} \}$, and $\{ \alpha_{5n} \}$. In this manuscript, we have considered the Johannsen metric with a possible non-vanishing $\alpha_{13}$ and set all other deformation parameters to zero. This is the most widely used choice in the literature of X-ray tests of the Kerr hypothesis.



\begin{thebibliography}{99}

\bibitem{Will:2014kxa}
C.~M.~Will,
Living Rev. Rel. \textbf{17}, 4 (2014)
doi:10.12942/lrr-2014-4
[arXiv:1403.7377 [gr-qc]].

\bibitem{Bambi:2015kza}
C.~Bambi,
Rev. Mod. Phys. \textbf{89}, 025001 (2017)
doi:10.1103/RevModPhys.89.025001
[arXiv:1509.03884 [gr-qc]].

\bibitem{Bambi:2017khi}
C.~Bambi,
{\it Black Holes: A Laboratory for Testing Strong Gravity},
(Springer Singapore, 2017),
ISBN 978-981-10-4524-0,
doi:10.1007/978-981-10-4524-0

\bibitem{Yagi:2016jml}
K.~Yagi and L.~C.~Stein,
Class. Quant. Grav. \textbf{33}, 054001 (2016)
doi:10.1088/0264-9381/33/5/054001
[arXiv:1602.02413 [gr-qc]].

\bibitem{Barack:2018yly}
L.~Barack, V.~Cardoso, S.~Nissanke, T.~P.~Sotiriou, A.~Askar, C.~Belczynski, G.~Bertone, E.~Bon, D.~Blas and R.~Brito, \textit{et al.}
Class. Quant. Grav. \textbf{36}, 143001 (2019)
doi:10.1088/1361-6382/ab0587
[arXiv:1806.05195 [gr-qc]].

\bibitem{LIGOScientific:2016lio}
B.~P.~Abbott \textit{et al.} [LIGO Scientific and Virgo],
Phys. Rev. Lett. \textbf{116}, 221101 (2016)
[erratum: Phys. Rev. Lett. \textbf{121}, 129902 (2018)]
doi:10.1103/PhysRevLett.116.221101
[arXiv:1602.03841 [gr-qc]].

\bibitem{Yunes:2016jcc}
N.~Yunes, K.~Yagi and F.~Pretorius,
Phys. Rev. D \textbf{94}, 084002 (2016)
doi:10.1103/PhysRevD.94.084002
[arXiv:1603.08955 [gr-qc]].

\bibitem{LIGOScientific:2020tif}
R.~Abbott \textit{et al.} [LIGO Scientific and Virgo],
Phys. Rev. D \textbf{103}, 122002 (2021)
doi:10.1103/PhysRevD.103.122002
[arXiv:2010.14529 [gr-qc]].

\bibitem{Shashank:2021giy}
S.~Shashank and C.~Bambi,
Phys. Rev. D \textbf{105}, 104004 (2022)
doi:10.1103/PhysRevD.105.104004
[arXiv:2112.05388 [gr-qc]].

\bibitem{Cao:2017kdq}
Z.~Cao, S.~Nampalliwar, C.~Bambi, T.~Dauser and J.~A.~Garcia,
Phys. Rev. Lett. \textbf{120}, 051101 (2018)
doi:10.1103/PhysRevLett.120.051101
[arXiv:1709.00219 [gr-qc]].

\bibitem{Tripathi:2018lhx}
A.~Tripathi, S.~Nampalliwar, A.~B.~Abdikamalov, D.~Ayzenberg, C.~Bambi, T.~Dauser, J.~A.~Garcia and A.~Marinucci,
Astrophys. J. \textbf{875}, 56 (2019)
doi:10.3847/1538-4357/ab0e7e
[arXiv:1811.08148 [gr-qc]].

\bibitem{Tripathi:2020dni}
A.~Tripathi, A.~B.~Abdikamalov, D.~Ayzenberg, C.~Bambi, V.~Grinberg and M.~Zhou,
Astrophys. J. \textbf{907}, 31 (2021)
doi:10.3847/1538-4357/abccbd
[arXiv:2010.13474 [astro-ph.HE]].

\bibitem{Tripathi:2020yts}
A.~Tripathi, Y.~Zhang, A.~B.~Abdikamalov, D.~Ayzenberg, C.~Bambi, J.~Jiang, H.~Liu and M.~Zhou,
Astrophys. J. \textbf{913}, 79 (2021)
doi:10.3847/1538-4357/abf6cd
[arXiv:2012.10669 [astro-ph.HE]].

\bibitem{Bambi:2019tjh}
C.~Bambi, K.~Freese, S.~Vagnozzi and L.~Visinelli,
Phys. Rev. D \textbf{100}, 044057 (2019)
doi:10.1103/PhysRevD.100.044057
[arXiv:1904.12983 [gr-qc]].

\bibitem{EventHorizonTelescope:2020qrl}
D.~Psaltis \textit{et al.} [Event Horizon Telescope],
Phys. Rev. Lett. \textbf{125}, 141104 (2020)
doi:10.1103/PhysRevLett.125.141104
[arXiv:2010.01055 [gr-qc]].

\bibitem{EventHorizonTelescope:2022xqj}
K.~Akiyama \textit{et al.} [Event Horizon Telescope],
Astrophys. J. Lett. \textbf{930}, L17 (2022)
doi:10.3847/2041-8213/ac6756

\bibitem{Vagnozzi:2022moj}
S.~Vagnozzi, R.~Roy, Y.~D.~Tsai, L.~Visinelli, M.~Afrin, A.~Allahyari, P.~Bambhaniya, D.~Dey, S.~G.~Ghosh and P.~S.~Joshi, \textit{et al.}
Class. Quant. Grav. \textbf{40}, 165007 (2023)
doi:10.1088/1361-6382/acd97b
[arXiv:2205.07787 [gr-qc]].

\bibitem{Kerr:1963ud}
R.~P.~Kerr,
Phys. Rev. Lett. \textbf{11}, 237-238 (1963)
doi:10.1103/PhysRevLett.11.237

\bibitem{Bambi:2014koa}
C.~Bambi, D.~Malafarina and N.~Tsukamoto,
Phys. Rev. D \textbf{89}, 127302 (2014)
doi:10.1103/PhysRevD.89.127302
[arXiv:1406.2181 [gr-qc]].

\bibitem{Bambi:2008hp}
C.~Bambi, A.~D.~Dolgov and A.~A.~Petrov,
JCAP \textbf{09}, 013 (2009)
doi:10.1088/1475-7516/2009/09/013
[arXiv:0806.3440 [astro-ph]].

\bibitem{Dvali:2011aa}
G.~Dvali and C.~Gomez,
Fortsch. Phys. \textbf{61}, 742-767 (2013)
doi:10.1002/prop.201300001
[arXiv:1112.3359 [hep-th]].

\bibitem{Herdeiro:2014goa}
C.~A.~R.~Herdeiro and E.~Radu,
Phys. Rev. Lett. \textbf{112}, 221101 (2014)
doi:10.1103/PhysRevLett.112.221101
[arXiv:1403.2757 [gr-qc]].

\bibitem{Giddings:2014ova}
S.~B.~Giddings,
Phys. Rev. D \textbf{90}, 124033 (2014)
doi:10.1103/PhysRevD.90.124033
[arXiv:1406.7001 [hep-th]].

\bibitem{NuSTAR:2013yza}
F.~A.~Harrison \textit{et al.} [NuSTAR],
Astrophys. J. \textbf{770}, 103 (2013)
doi:10.1088/0004-637X/770/2/103
[arXiv:1301.7307 [astro-ph.IM]].

\bibitem{Bambi:2016wdn}
C.~Bambi, L.~Modesto and L.~Rachwa\l{},
JCAP \textbf{05}, 003 (2017)
doi:10.1088/1475-7516/2017/05/003
[arXiv:1611.00865 [gr-qc]].

\bibitem{Sen:1992ua}
A.~Sen,
Phys. Rev. Lett. \textbf{69}, 1006-1009 (1992)
doi:10.1103/PhysRevLett.69.1006
[arXiv:hep-th/9204046 [hep-th]].

\bibitem{Johannsen:2013szh}
T.~Johannsen,
Phys. Rev. D \textbf{88}, 044002 (2013)
doi:10.1103/PhysRevD.88.044002
[arXiv:1501.02809 [gr-qc]].

\bibitem{Zhang:2021ymo}
Z.~Zhang, H.~Liu, A.~B.~Abdikamalov, D.~Ayzenberg, C.~Bambi and M.~Zhou,
Astrophys. J. \textbf{924}, 72 (2022)
doi:10.3847/1538-4357/ac350e
[arXiv:2106.03086 [astro-ph.HE]].

\bibitem{Tripathi:2021rqs}
A.~Tripathi, A.~B.~Abdikamalov, D.~Ayzenberg, C.~Bambi, V.~Grinberg, H.~Liu and M.~Zhou,
JCAP \textbf{01}, 019 (2022)
doi:10.1088/1475-7516/2022/01/019
[arXiv:2106.10982 [astro-ph.HE]].

\bibitem{Bambi:2022dtw}
C.~Bambi,
[arXiv:2210.05322 [gr-qc]].

\bibitem{Psaltis:2020ctj}
D.~Psaltis, C.~Talbot, E.~Payne and I.~Mandel,
Phys. Rev. D \textbf{103}, 104036 (2021)
doi:10.1103/PhysRevD.103.104036
[arXiv:2012.02117 [gr-qc]].

\bibitem{Bambi:2020jpe}
C.~Bambi, L.~W.~Brenneman, T.~Dauser, J.~A.~Garcia, V.~Grinberg, A.~Ingram, J.~Jiang, H.~Liu, A.~M.~Lohfink and A.~Marinucci, \textit{et al.}
Space Sci. Rev. \textbf{217}, 65 (2021)
doi:10.1007/s11214-021-00841-8
[arXiv:2011.04792 [astro-ph.HE]].

\bibitem{Bambi:2021chr}
C.~Bambi,
Arab. J. Math. \textbf{11}, 81-90 (2022)
doi:10.1007/s40065-021-00336-y
[arXiv:2106.04084 [gr-qc]].

\bibitem{Bambi:2012tg}
C.~Bambi,
Astrophys. J. \textbf{761}, 174 (2012)
doi:10.1088/0004-637X/761/2/174
[arXiv:1210.5679 [gr-qc]].

\bibitem{Ross:2005dm}
R.~R.~Ross and A.~C.~Fabian,
Mon. Not. Roy. Astron. Soc. \textbf{358}, 211-216 (2005)
doi:10.1111/j.1365-2966.2005.08797.x
[arXiv:astro-ph/0501116 [astro-ph]].

\bibitem{Garcia:2010iz}
J.~Garcia and T.~Kallman,
Astrophys. J. \textbf{718}, 695 (2010)
doi:10.1088/0004-637X/718/2/695
[arXiv:1006.0485 [astro-ph.HE]].

\bibitem{Bambi:2016sac}
C.~Bambi, A.~Cardenas-Avendano, T.~Dauser, J.~A.~Garcia and S.~Nampalliwar,
Astrophys. J. \textbf{842}, 76 (2017)
doi:10.3847/1538-4357/aa74c0
[arXiv:1607.00596 [gr-qc]].

\bibitem{Abdikamalov:2019yrr}
A.~B.~Abdikamalov, D.~Ayzenberg, C.~Bambi, T.~Dauser, J.~A.~Garcia and S.~Nampalliwar,
Astrophys. J. \textbf{878}, 91 (2019)
doi:10.3847/1538-4357/ab1f89
[arXiv:1902.09665 [gr-qc]].

\bibitem{Abdikamalov:2020oci}
A.~B.~Abdikamalov, D.~Ayzenberg, C.~Bambi, T.~Dauser, J.~A.~Garcia, S.~Nampalliwar, A.~Tripathi and M.~Zhou,
Astrophys. J. \textbf{899}, 80 (2020)
doi:10.3847/1538-4357/aba625
[arXiv:2003.09663 [astro-ph.HE]].

\bibitem{Dauser:2013xv}
T.~Dauser, J.~Garcia, J.~Wilms, M.~Bock, L.~W.~Brenneman, M.~Falanga, K.~Fukumura and C.~S.~Reynolds,
Mon. Not. Roy. Astron. Soc. \textbf{430}, 1694 (2013)
doi:10.1093/mnras/sts710
[arXiv:1301.4922 [astro-ph.HE]].

\bibitem{Garcia:2013oma}
J.~Garcia, T.~Dauser, C.~S.~Reynolds, T.~R.~Kallman, J.~E.~McClintock, J.~Wilms and W.~Eikmann,
Astrophys. J. \textbf{768}, 146 (2013)
doi:10.1088/0004-637X/768/2/146
[arXiv:1303.2112 [astro-ph.HE]].

\bibitem{Garcia:2013lxa}
J.~Garc\'\i{}a, T.~Dauser, A.~Lohfink, T.~R.~Kallman, J.~Steiner, J.~E.~McClintock, L.~Brenneman, J.~Wilms, W.~Eikmann and C.~S.~Reynolds, \textit{et al.}
Astrophys. J. \textbf{782}, 76 (2014)
doi:10.1088/0004-637X/782/2/76
[arXiv:1312.3231 [astro-ph.HE]].

\bibitem{Bambi:2023czx}
C.~Bambi, A.~B.~Abdikamalov, H.~Liu, S.~Riaz, S.~Shashank and M.~Zhou,
[arXiv:2307.12755 [astro-ph.IM]].

\bibitem{Zhou:2018bxk}
M.~Zhou, Z.~Cao, A.~Abdikamalov, D.~Ayzenberg, C.~Bambi, L.~Modesto and S.~Nampalliwar,
Phys. Rev. D \textbf{98}, 024007 (2018)
doi:10.1103/PhysRevD.98.024007
[arXiv:1803.07849 [gr-qc]].

\bibitem{Zhou:2019hqk}
M.~Zhou, A.~B.~Abdikamalov, D.~Ayzenberg, C.~Bambi, L.~Modesto, S.~Nampalliwar and Y.~Xu,
EPL \textbf{125}, 30002 (2019)
doi:10.1209/0295-5075/125/30002
[arXiv:2003.03738 [gr-qc]].

\bibitem{Tripathi:2021rwb}
A.~Tripathi, B.~Zhou, A.~B.~Abdikamalov, D.~Ayzenberg and C.~Bambi,
JCAP \textbf{07}, 002 (2021)
doi:10.1088/1475-7516/2021/07/002
[arXiv:2103.07593 [astro-ph.HE]].

\bibitem{Wilms:2000ez}
J.~Wilms, A.~Allen and R.~McCray,
Astrophys. J. \textbf{542}, 914-924 (2000)
doi:10.1086/317016
[arXiv:astro-ph/0008425 [astro-ph]].

\bibitem{Riaz:2019kat}
S.~Riaz, D.~Ayzenberg, C.~Bambi and S.~Nampalliwar,
Astrophys. J. \textbf{895}, 61 (2020)
doi:10.3847/1538-4357/ab89ab
[arXiv:1911.06605 [astro-ph.HE]].

\bibitem{eXTP:2016rzs}
S.~N.~Zhang \textit{et al.} [eXTP],
Proc. SPIE Int. Soc. Opt. Eng. \textbf{9905}, 99051Q (2016)
doi:10.1117/12.2232034
[arXiv:1607.08823 [astro-ph.IM]].

\bibitem{Nandra:2013jka}
K.~Nandra, D.~Barret, X.~Barcons, A.~Fabian, J.~W.~d.~Herder, L.~Piro, M.~Watson, C.~Adami, J.~Aird and J.~M.~Afonso, \textit{et al.}
[arXiv:1306.2307 [astro-ph.HE]].


\end{thebibliography}
\end{document}